\documentclass[aps,pre,twocolumn,superscriptaddress]{revtex4-2}

\usepackage{amsmath}
\usepackage{amssymb}
\usepackage{braket}
\usepackage{graphicx}
\usepackage{color}
\usepackage{mathtools}
\usepackage{enumitem}

\graphicspath{{Figures/}}

\newcommand{\zero}{\mathtt{0}}
\newcommand{\one}{\mathtt{1}}

\begin{document}
	
\title{Lasing and counter-lasing phase transitions in a cavity QED system}

\author{Kevin C. Stitely}
\email{kevin.stitely@auckland.ac.nz}
\affiliation{Dodd-Walls Centre for Photonic and Quantum Technologies, New Zealand}
\affiliation{Department of Physics, University of Auckland, New Zealand}
\affiliation{Department of Mathematics, University of Auckland, New Zealand}

\author{Andrus Giraldo}
\affiliation{Dodd-Walls Centre for Photonic and Quantum Technologies, New Zealand}
\affiliation{Department of Mathematics, University of Auckland, New Zealand}

\author{Bernd Krauskopf}
\affiliation{Dodd-Walls Centre for Photonic and Quantum Technologies, New Zealand}
\affiliation{Department of Mathematics, University of Auckland, New Zealand}

\author{Scott Parkins}
\affiliation{Dodd-Walls Centre for Photonic and Quantum Technologies, New Zealand}
\affiliation{Department of Physics, University of Auckland, New Zealand}

\date{\today}

\begin{abstract}
	We study the effect of spontaneous emission and incoherent atomic pumping on the nonlinear semiclassical dynamics of the unbalanced Dicke model -- a generalization of the Dicke model that features independent coupling strengths for the co- and counter-rotating interaction terms. As well as the ubiquitous superradiant behavior the Dicke model is well-known for, the addition of spontaneous emission combined with the presence of strong counter-rotating terms creates laser-like behavior termed \emph{counter-lasing}. These states appear in the semiclassical model as stable periodic orbits. We perform a comprehensive dynamical analysis of the appearance of counter-lasing in the unbalanced Dicke model subject to strong cavity dissipation, such that the cavity field can be adiabatically eliminated to yield an effective Lipkin-Meshkov-Glick (LMG) model. If the coupling strength of the co-rotating interactions is small, then the counter-lasing phase appears via a Hopf bifurcation of the de-excited state. We find that if the rate of spontaneous emission is small, this can lead to resurgent superradiant pulses. However, if the co-rotating coupling is larger, then the counter-lasing phase must emerge via the steady-state superradiant phase. Such a transition is the result of the competition of the coherent and incoherent processes that drive superradiance and counter-lasing, respectively. We observe a surprisingly complex transition between the two, associated with the formation of a chaotic attractor over a thin transitional parameter region. 
\end{abstract}
	
\maketitle

%
%

\section{Introduction}
Studies of the dynamics associated with models of trapped ultracold atomic gases and ion arrays have recently made tremendous progress towards both experimentally realizing and theoretically modelling systems that present a broad variety of complex dynamics. Increased technological and computational capabilities have made experimental and numerical investigation of many-body quantum systems now tractable \cite{muniz_exploring_2020,frisk_kockum_ultrastrong_2019}. Apart from being of importance to fundamental theoretical physics, such systems are of interest for their applications in the fields of metrology, developing state-of-the-art atomic clocks and other measurement devices \cite{ludlow_optical_2015,qu_spin_2019,jiang_making_2011}, and quantum computation and information processing \cite{haffner_quantum_2008,daley_quantum_2011}. 

One such model that has been garnering much interest of late is the Dicke model \cite{dicke_coherence_1954}, which describes the interaction of an ensemble of atoms confined to an optical cavity with a single electromagnetic field mode of that cavity. As later shown by Hepp and Lieb, this model is now renowned for a quantum phase transition to superradiance, where the emission of light by the atomic ensemble is enhanced by maintaining atomic coherence \cite{hepp_equilibrium_1973,hepp_superradiant_1973,kirton_introduction_2019}. Although difficult to realize in practice with two-level atoms, it is possible to engineer \emph{effective} Dicke models, which present superradiant dynamics in a wide variety of arrangements. Examples are Bose--Einstein condensates \cite{baumann_dicke_2010,baumann_exploring_2011,hamner_dicke-type_2014,klinder_dynamical_2015}, trapped-ion arrays \cite{safavi-naini_verification_2018}, and cavity quantum electrodynamics (QED) with multilevel atoms driven by Raman transitions \cite{dimer_proposed_2007,zhiqiang_nonequilibrium_2017,zhang_dicke-model_2018}, which have provoked much interest recently in studying their dynamics \cite{emary_chaos_2003,emary_quantum_2003,bhaseen_dynamics_2012,keeling_collective_2010,grimsmo_dissipative_2013,soriente_dissipation-induced_2018,stitely_nonlinear_2020}.

\begin{figure}[b]
	\centering
	\includegraphics[width=6.5cm]{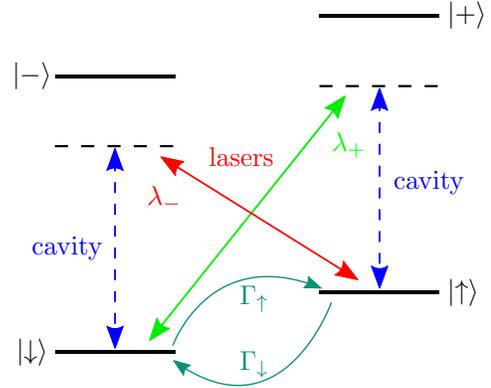}
	\caption{\label{fig:model} Schematic of the atomic energy level configuration with the two ground states $\ket{\downarrow}$ and $\ket{\uparrow}$, and the two excited states $\ket{-}$ and $\ket{+}$. Two lasers are driving two Raman transitions between the ground and excited states, which give rise to two coupling strengths $\lambda_{-}$ and $\lambda_{+}$, corresponding to the co-rotating and counter-rotating interactions in the Hamiltonian (\ref{eqn:DickeHamiltonian}), respectively. In addition, spontaneous emission and incoherent pumping between the two ground states may occur at rates $\Gamma_\downarrow$ and $\Gamma_\uparrow$.}
\end{figure}

We focus here particularly on the cavity QED realizations of the Dicke model. An effective Dicke model can be produced in this case by applying lasers stimulating two Raman transitions between two atomic ground states of an ensemble of four-level atoms. The scheme is illustrated in Fig. \ref{fig:model}. There are two ground states, $\ket{\downarrow}$ and $\ket{\uparrow}$, and two highly detuned excited states, $\ket{-}$ and $\ket{+}$. Two lasers drive the transitions $\ket{\downarrow}\leftrightarrow\ket{+}$ and $\ket{\uparrow}\leftrightarrow\ket{-}$, while the cavity mode couples the excited and ground states through the transitions $\ket{\downarrow}\leftrightarrow\ket{-}$ and $\ket{\uparrow}\leftrightarrow\ket{+}$. If the excited state detunings are large, then the two atomic excited states can be adiabatically eliminated such that only the dynamics between the two ground states need to be considered. This model must be considered as an open quantum system, as photon loss through cavity decay is unavoidable. The resulting superradiant quantum phase transition is mediated by the atom-light coupling strength. If the coupling is sufficiently high, the system transitions from the so-called normal phase, where the atoms are in their (nominal) ground states $\ket{\downarrow}$ and the light field is in the vacuum state, to the superradiant phase, where the atoms and the cavity feature macroscopic excitation in the steady state. 

This effective realization presents an interesting generalization of the Dicke model if the lasers driving the two Raman transitions are not of the same strength or frequency. In the theoretical model, the possibility of having two driving lasers with different frequencies and electric field amplitudes leads to different coupling strengths for the co-rotating and counter-rotating terms of the interaction Hamiltonian, leading to the \emph{unbalanced Dicke model} -- also referred to in the literature as the anisotropic Dicke model \cite{buijsman_nonergodicity_2017}, the generalized Jaynes-Cummings-Rabi model \cite{ricardo_dissipative_2018}, and the interpolating Dicke-Tavis-Cummings model \cite{soriente_dissipation-induced_2018}. This generalization provides a surprisingly diverse range of dynamical behavior, including normal-superradiant phase coexistence \cite{keeling_collective_2010,soriente_dissipation-induced_2018}, hysteresis \cite{stitely_superradiant_2020}, superradiant oscillations \cite{zhiqiang_nonequilibrium_2017}, and chaos \cite{stitely_nonlinear_2020}.

Among this behavior is the \emph{pole-flip transition} \cite{stitely_nonlinear_2020}, which describes a transition between the two types of atomic configurations in the normal phase: either all spins down or all spins up. This transition was observed experimentally in Ref. \cite{zhiqiang_nonequilibrium_2017}. However, results indicated that the transition threshold was not well resolved compared to the expectations provided by theory. A subsequent theoretical analysis using a fermionic Keldysh path integral formulation \cite{shchadilova_fermionic_2020} found that the discrepancy can be explained by considering atomic decay channels that break the conservation of the collective spin length. The inclusion of atomic dissipation destabilizes the inverted normal phase (where all spins are up) and can produce a type of lasing state termed \emph{counter-lasing} \cite{shchadilova_fermionic_2020,kirton_introduction_2019}, which is ``fed" by the counter-rotating interactions; this is in contrast to usual lasing which is fed by co-rotating interactions. Other effects brought about by the inclusion of both individual and collective atomic dissipation have been investigated in Refs. \cite{gelhausen_many-body_2017,gelhausen_dissipative_2018,kirton_suppressing_2017}.

With the addition of an incoherent pump exciting the spins, regular lasing states can be achieved as well \cite{kirton_superradiant_2018}. It was found that, in a semiclassical treatment, the establishment of the lasing phase occurs via a Hopf bifurcation leading to the appearance of a stable periodic orbit. 

In this work, we take a bifurcation theory approach to studying the emergence of lasing and counter-lasing within a semiclassical treatment. We first study the emergence of lasing and counter-lasing as a modification of the pole-flip transition, which occurs in the absence of atomic dissipation or pumping. We find that the transition is most naturally described in the bad-cavity or dissipative limit, where the cavity field mode can be adiabatically eliminated to yield a spin-only description of the relevant dynamics. For this model, we perform a detailed analysis of the periodic orbits describing lasing and counter-lasing states, resulting from Hopf bifurcations. We then investigate the transition from superradiance to the counter-lasing phase, which we find occurs via a small transitional parameter region across which infinitely many (global) bifurcations occur, leading to the emergence of chaotic dynamics. 

In the counter-lasing phase the spontaneous emission rate acts as a \emph{pump} for the lasing process, in conjunction with the coherent counter-rotating interactions between the atomic ensemble and the light field. The superradiant phase is driven by both of the coherent co- and counter-rotating interactions, while the counter-lasing phase is seeded by incoherent atomic dissipation and the counter-rotating interactions. As such, the transition from superradiance to the counter-lasing phase allows us to explore the competition between coherent and incoherent processes of the model. This interplay results in a remarkably complex transition, which we probe in detail by analyzing the symbolic dynamics of the model via a topological quantity called the kneading invariant.

The paper is organized as follows. In Sect. II we introduce the model and its semiclassical approximation which we will study throughout the rest of the paper. In Sect. III we study the transition from the normal phase to lasing and counter-lasing states in the ``strong dissipative limit" where the cavity decay rate is large in comparision to the effective cavity resonance frequency.  In Sect. IV we consider the transition from superradiance to the counter-lasing phase, which reveals an abundance of complex chaotic dynamics within a small transitional region of parameter space. Here, we present a parameter sweeping of a kneading invariant to showcase the intricate organization of global bifurcations. Our conclusions are presented in Sect. V.

%
%

\section{Model}

We now introduce the unbalanced Dicke model with photon loss, atomic decay and pumping. We consider the model in the dissipative limit for large photon decay rates, where the field mode can be adiabatically eliminated to yield another effective Hamiltonian. We then introduce the semiclassical limit, given by a system of nonlinear ordinary differential equations.

\subsection{Quantum Mechanical Model}

The unbalanced Dicke model considers an ensemble of $N$ two-level atoms interacting with a single mode of the electromagnetic field of an optical cavity. The Hamiltonian (with $\hbar = 1$) is
\begin{align}\label{eqn:DickeHamiltonian}
\hat{H} = \ &\omega\hat{a}^{\dagger}\hat{a} + \omega_0\hat{J}_{z} + \frac{\lambda_{-}}{\sqrt{N}}\left( \hat{a}\hat{J}_{+} + \hat{a}^{\dagger}\hat{J}_{-} \right) \nonumber \\
& + \frac{\lambda_{+}}{\sqrt{N}}\left( \hat{a}\hat{J}_{-} + \hat{a}^\dagger\hat{J}_{+} \right),
\end{align}
where $\hat{a}$ is the annihilation operator of the cavity field mode, and
\begin{equation}
	\hat{J}_{\pm,x,y,z} = \sum_{j=1}^{N} \hat{\sigma}^{(j)}_{\pm,x,y,z}
\end{equation}
are collective angular momentum operators; here $\hat{\sigma}^{(j)}_{\pm,x,y,z}$ are the spin-$\frac{1}{2}$ Pauli operators for the $j$th atom. These operators obey the commutation relations $[\hat{\sigma}^{(j)}_{m},\hat{\sigma}^{(j)}_{n}] = i\varepsilon_{mn\ell}\hat{\sigma}^{(j)}_{\ell}$ (following the Einstein summation convention), where the indices $m$, $n$, $\ell$ take on the values $x$, $y$, $z$ and $\varepsilon_{mn\ell}$ is the Levi-Civita symbol. Here, $\hat{\sigma}^{(j)}_{\pm} = \hat{\sigma}^{(j)}_{x} \pm i\hat{\sigma}^{(j)}_{y}$ are the angular momentum raising and lowering operators. The effective frequency of the cavity mode is $\omega$, and the effective frequency splitting of the atomic energy levels is $\omega_0$. The co-rotating and counter-rotating coupling strengths are $\lambda_{-}$ and $\lambda_{+}$, respectively.

With the addition of cavity decay, and incoherent (individual) atomic dissipation and pumping, the quantum dynamics are governed by the master equation
\begin{equation}\label{eqn:fullmastereqn}
	\frac{d\hat{\rho}}{dt} = -i[\hat{H},\hat{\rho}] + \kappa\mathcal{D}\{\hat{a}\}\hat{\rho} + \sum_{j=1}^{N}\Gamma_{\downarrow}\mathcal{D}\{\hat{\sigma}^{(j)}_{-}\}\hat{\rho} + \Gamma_{\uparrow}\mathcal{D}\{\hat{\sigma}^{(j)}_{+}\}\hat{\rho},
\end{equation}
where $\hat{\rho}$ is the density operator, $\kappa$ is the cavity field decay rate and $\Gamma_{\downarrow}$ and $\Gamma_{\uparrow}$ are the atomic dissipation and pumping rates, respectively (see Fig. \ref{fig:model}). The operator
\begin{equation}
	\mathcal{D}\{\hat{X}\}\hat{\rho} = 2\hat{X}\hat{\rho}\hat{X}^\dagger - \hat{X}^\dagger\hat{X}\hat{\rho} - \hat{\rho}\hat{X}^\dagger\hat{X}
\end{equation}
is the standard Lindblad dissipator for a jump operator $\hat{X}$. Equation (\ref{eqn:fullmastereqn}) is invariant under the exchange
\begin{equation}
\hat{a} \leftrightarrow -\hat{a},\ \hat{a}^\dagger \leftrightarrow -\hat{a}^\dagger,\ \hat{\sigma}_{-}^{(j)} \leftrightarrow -\hat{\sigma}_{-}^{(j)},\ \hat{\sigma}_{+}^{(j)} \leftrightarrow -\hat{\sigma}_{+}^{(j)},
\end{equation}
which gives rise to a $\mathbb{Z}_{2}$ (parity) symmetry.

In either the dissipative limit $\kappa \gg \omega,\omega_0,\lambda_{\pm}$ \cite{morrison_dissipation-driven_2008} or the dispersive limit $\omega \gg \kappa,\omega_0,\lambda_{\pm}$ \cite{masson_cavity_2017,morrison_collective_2008} excitations in the cavity mode are short-lived. In both limits, the timescales for the atomic and photonic dynamics can be separated and the field mode adiabatically eliminated. That is, the cavity field is treated as mediating (long-range) atomic interactions. The resulting Hamiltonian
\begin{equation}\label{eqn:HamiltonianLMG}
	\hat{H}_{\mathrm{LMG}} = \omega_0'\hat{J}_{z} - \frac{\xi}{N}\left[ (\lambda_{-} + \lambda_{+})^2\hat{J}_{x}^2 + (\lambda_{-} - \lambda_{+})^2\hat{J}_{y}^2 \right]
\end{equation}
is a generalized form of a Lipkin-Meshkov-Glick (LMG) model  \cite{lipkin_validity_1965} describing the spin interactions. Including dissipative and incoherent pumping terms introduces the master equation
\begin{align}\label{eqn:mastereqn}
	\frac{d\hat{\rho}}{dt} =\ & -i[\hat{H}_{\mathrm{LMG}},\hat{\rho}] + \frac{\eta}{N}\mathcal{D}\{\lambda_{-}\hat{J}_{-} + \lambda_{+}\hat{J}_{+}\}\hat{\rho} \nonumber \\
	& + \sum_{j=1}^{N}\Gamma_{\downarrow}\mathcal{D}\{\hat{\sigma}_{-}^{(j)}\}\hat{\rho} + \Gamma_{\uparrow}\mathcal{D}\{\hat{\sigma}_{+}^{(j)}\}\hat{\rho},
\end{align}
where 
\begin{equation}\label{eqn:parameters}
\omega_0' = \omega_0 - \frac{\xi}{N}\left( \lambda_{-}^2 - \lambda_{+}^2 \right),\ \xi = \frac{\omega}{\kappa^2 + \omega^2},\ \eta = \frac{\kappa}{\kappa^2 + \omega^2}.
\end{equation}
Equation (\ref{eqn:mastereqn}) constitutes the quantum mechanical generalized LMG model that we consider in this paper.

\subsection{Semiclassical Model}

We take a semiclassical approach valid for large numbers of atoms $N\rightarrow\infty$, where the expectations of operator products factorize, that is, quantum fluctuations are negligible. Moreover, the atoms couple uniformly to the cavity mode and feature identical atomic decay and pumping rates, so that the density operator site-decouples as $\hat{\rho} = \bigotimes_{j=1}^{N}\hat{\rho}_{j}$, where $\hat{\rho}_{j}$ is the density operator for the $j$th atom \cite{gelhausen_many-body_2017,kirton_suppressing_2017}. Hence, the dynamics described by the master equation (\ref{eqn:mastereqn}) can be represented as a closed system of nonlinear differential equations. Defining the variables
\begin{equation}
\beta = \frac{\braket{\hat{J}_-}}{N} = b_{x} - ib_y \in\mathbb{C},\ \gamma = \frac{\braket{\hat{J}_{z}}}{N}\in\mathbb{R},
\end{equation}
the semiclassical equations of motion are then
\begin{align}
	\frac{d\beta}{dt} ={}& -i\omega_0\beta - 2\eta(\lambda_{+}^2 - \lambda_{-}^2)\beta\gamma - 2i\xi(\lambda_{+}^2 + \lambda_{-}^2)\beta\gamma \nonumber \\
	& - 4i\xi\lambda_{-}\lambda_{+}\beta^* \gamma - (\Gamma_\uparrow + \Gamma_\downarrow)\beta, \label{eqn:ODEs1} \\
	\frac{d\gamma}{dt} ={}& 2\eta(\lambda_{+}^2 - \lambda_{-}^2)\beta^* \beta + 2i\xi\lambda_{-}\lambda_{+}[(\beta^*)^2 - \beta^2] \nonumber \\
	& -2\Gamma_\downarrow \left( \frac{1}{2} + \gamma \right) + 2\Gamma_\uparrow\left(\frac{1}{2} - \gamma  \right), \label{eqn:ODEs2}
\end{align}
which constitute the semiclassical generalized LMG model that we consider throughout the paper. The nonlinear dynamics described by these equations is the central focus of this paper. They form a dynamical system with a three-dimensional phase space $(b_{x},b_{y},\gamma)\in\mathbb{R}^{3}$ since $\beta$ is complex. As is the case for Eq.~(\ref{eqn:fullmastereqn}), Eqs.~(\ref{eqn:ODEs1})--(\ref{eqn:ODEs2}) feature $\mathbb{Z}_{2}$ (parity) symmetry given by the equivariance \cite{golubitskii_singularities_1985} under the transformation
\begin{equation}\label{eqn:symmetryExchange}
\mathcal{T}: \beta \mapsto -\beta,
\end{equation}
which corresponds to a rotation by an angle $\pi$ about the $\gamma$-axis in $(b_{x},b_{y},\gamma)$-space. This symmetry produces trajectories that either come in pairs, with the symmetric counterpart related by $\mathcal{T}$, or are themselves symmetric under $\mathcal{T}$.

When the atomic pumping and damping terms vanish (i.e., $\Gamma_\uparrow = \Gamma_\downarrow=0$) angular momentum is conserved and trajectories are confined onto the Bloch sphere defined by
\begin{equation}\label{eqn:conservation}
b_{x}^2 + b_{y}^2 + \gamma^2 = R^2.
\end{equation}
Here $0<R\leq1/2$ is the radius of the Bloch sphere, determined by the initial atomic polarization. The maximal atomic polarization $R=1/2$ is determined by the total spin length $N/2$ of $N$ spin-$\frac{1}{2}$ atoms. Hence, the phase space of Eqs. (\ref{eqn:ODEs1})--(\ref{eqn:ODEs2}) in the absence of atomic dissipation ($\Gamma_\downarrow = 0$) and pumping ($\Gamma_\uparrow = 0$) is the sphere $\mathbb{S}^2(R)$ of radius $R$.

%
%

\section{Strong Dissipative Limit}

Analysis of the semiclassical dynamics of the unbalanced Dicke model without atomic dissipation or pumping indicates two distinct types of pole-flip transition: the normal and superradiant cases \cite{stitely_nonlinear_2020}. The normal pole-flip transition describes the transition from the spin-down normal phase to the inverted (spin-up) normal phase, while the superradiant pole-flip transition describes the transition from superradiance to the inverted normal phase. In this section we focus on the simpler normal pole-flip transition. To this end, we consider the \emph{strong dissipative limit} $\xi=0$ ($\kappa \gg \omega$).

In this regime, Eqs.~(\ref{eqn:ODEs1})--(\ref{eqn:ODEs2}) are most conveniently represented in spherical coordinates
\begin{align}
	b_{x} &= r\sin(\theta)\cos(\phi), \\
	b_{y} &= r\sin(\theta)\sin(\phi), \\
	\gamma &= r\cos(\theta),
\end{align}
with radius $r\geq 0$, polar angle $\theta\in [0,\pi]$ (measured from the positive $\gamma$-axis), and azimuthal angle $\phi\in [0,2\pi)$ (measured from the positive $b_{x}$-axis). This leads to the set of ordinary differential equations:
\begin{align}
	\frac{dr}{dt} &= \delta\cos(\theta) - \sigma r \cos^{2}(\theta) - \sigma r, \label{eqn:SphereODEs1} \\
	\frac{d\theta}{dt} &= -\left[ \mu r - \sigma \cos(\theta) + \frac{\delta}{r} \right]\sin(\theta),  \label{eqn:SphereODEs2} \\
	\frac{d\phi}{dt} &= \omega_0  \label{eqn:SphereODEs3},
\end{align}
where we have introduced the parameters
\begin{equation}
\mu = 2\eta(\lambda_{+}^2 - \lambda_{-}^2),\ \ \sigma = \Gamma_\uparrow + \Gamma_\downarrow,\ \ \delta = \Gamma_\uparrow - \Gamma_\downarrow.
\end{equation}

\begin{figure}[t]
	\centering
	\includegraphics[width=7cm]{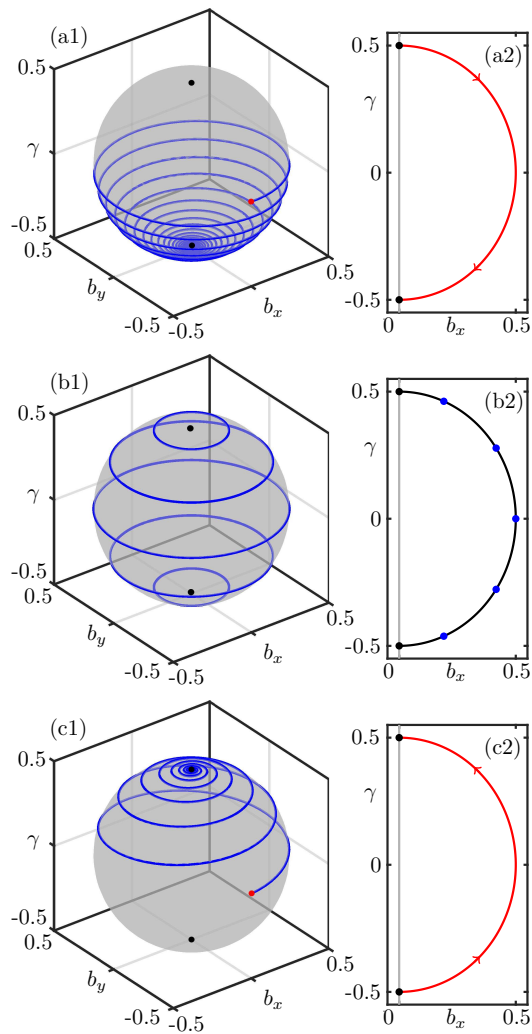}
	\caption{\label{fig:NormalPoleFlip}Normal pole-flip transition on the Bloch sphere in the generalized semiclassical LMG model in the strong dissipative limit, given by Eqs.~(\ref{eqn:ODEs1})--(\ref{eqn:ODEs2}). Panel (a1) shows a trajectory on the Bloch sphere in the spin-down normal phase and panel (a2) the trajectory in the $(b_{x},\gamma)$-half-plane of the rotating frame. Panel (b1) illustrates a selection of periodic orbits that appear during the normal pole-flip transition for $\lambda_{+} = 0.5$, where in panel (b2) these periodic orbits appear as equilibria in the rotating frame. In panel (c1) the system is now in the spin-up normal phase for $\lambda_{+} = 0.8$, and its rotating frame trajectory is shown in panel (c2). Other parameters for all plots are $\kappa = 5$, $\omega = 0$, $\omega_0 = 1$, $\lambda_{-} = 0.5$, $\Gamma_\uparrow = \Gamma_\downarrow = 0$.}
\end{figure}

In these spherical coordinates Eq. (\ref{eqn:SphereODEs3}) for $\phi$ decouples from Eqs.~(\ref{eqn:SphereODEs1}) and (\ref{eqn:SphereODEs2}) for $r$ and $\theta$. Furthermore, $d\phi/dt$ is constant, meaning that system (\ref{eqn:ODEs1})--(\ref{eqn:ODEs2}) features a constant rotation about the $\gamma$-axis in $(b_x,b_y,\gamma)$-space at the frequency $\omega_0$. Thus, in the strong dissipative limit, the semiclassical equations of motion are invariant under rotations of any angle about the $\gamma$-axis. That is, Eqs.~(\ref{eqn:ODEs1})--(\ref{eqn:ODEs2}) have the continuous $U(1)\cong\mathbb{S}^{1}$ symmetry, given by the equivariance under the map
\begin{equation}
\mathcal{S}: \beta \mapsto e^{i\phi}\beta,
\end{equation}
which contains the discrete $\mathbb{Z}_{2}$ symmetry as the special case $\phi = \pi$. This has the effect that the overall dynamics of Eqs.~(\ref{eqn:ODEs1})--(\ref{eqn:ODEs2}) is given by the dynamics in the $(r,\theta)$-half-plane with $r\geq 0$, subject to a constant rotation about the $\gamma$-axis at frequency $\omega_0$.

\subsection{Normal Pole-Flip Transition}\label{SubSect:NormalPoleFlip}

In the case of zero atomic damping or pumping, $\Gamma_{\downarrow} = \Gamma_{\uparrow} = \sigma = \delta = 0$, the dynamics become even simpler. Here, angular momentum is now conserved [Eq.~(\ref{eqn:conservation})], and the equations of motion in spherical coordinates become
\begin{align}
	\frac{dr}{dt} &= 0, \label{eqn:NormalPoleFlip1} \\
	\frac{d\theta}{dt} &= -\mu r\sin(\theta), \label{eqn:NormalPoleFlip2}\\
	\frac{d\phi}{dt} &= \omega_0. \label{eqn:NormalPoleFlip3}
\end{align}
Equation (\ref{eqn:NormalPoleFlip1}) means the dynamics are confined to invariant spherical surfaces parameterized by $R$ with $0<R\leq 1/2$. Additionally, from Eq. (\ref{eqn:ODEs2}) it follows that the entire $\gamma$-axis is a one-dimensional \emph{manifold of equilibria}, since every value of $\gamma$ for $\beta=0$ is an equilibrium point. This manifold's intersection with one of the spheres creates the two (spin-down and spin-up) normal phase equilibria at the South ($\theta = \pi$) and North ($\theta=0$) poles of the Bloch sphere, respectively. The dynamics on the sphere are essentially one-dimensional, determined by Eq. (\ref{eqn:NormalPoleFlip2}).

Since $\sin(\theta) > 0$ for $0\leq\theta\leq\pi$, the sign of $\mu$ then determines which equilibrium is approached, and thus determines the phase. If $\mu<0$ (i.e., $\lambda_{-}>\lambda_{+}$) then $\theta(t)\rightarrow \pi$ as $t\rightarrow\infty$ and the system is in the spin-down normal phase with $\gamma\rightarrow -\frac{1}{2}$. Hence, trajectories on the Bloch sphere approach the South pole in a rotational fashion. This is illustrated on the Bloch sphere in Fig. \ref{fig:NormalPoleFlip}(a1) for an initial condition on the equator in order to highlight the direction of the flow. It is illustrative here to consider the dynamics without the constant frequency rotation of $\phi$, that is, consider only Eqs. (\ref{eqn:SphereODEs1})--(\ref{eqn:SphereODEs2}); the corresponding trajectory in the rotating frame in the $(b_{x},\gamma)$-half-plane is shown in Fig. \ref{fig:NormalPoleFlip}(a2). As the atoms become de-excited during their traversal of the Bloch sphere, they rapidly release a pulse of photons in a manner referred to as \emph{transient} or \emph{Dicke} superradiance \cite{gross_superradiance:_1982}.

In contrast, if $\mu > 0$ (i.e., $\lambda_{+}>\lambda_{-}$) then $\theta(t)\rightarrow 0$ as $t\rightarrow\infty$ and the system is in the inverted spin-up normal phase; see Fig. \ref{fig:NormalPoleFlip}(c1) and the flow in the rotating frame in Fig. \ref{fig:NormalPoleFlip}(c2). Here, trajectories approach the equilibrium point at the North pole of the Bloch sphere. 

The transition between these two cases for $\mu = 0$ is the pole-flip transition. The dynamics here are highly degenerate, as $d\theta/dt = 0$. Then, subject to the constant rotation of the angle $\phi$, the Bloch sphere (again for a fixed radius $0<R\leq 1/2$) is foliated by infinitely many circular periodic orbits, which lie in planes of constant $\gamma$. This is illustrated in Fig. \ref{fig:NormalPoleFlip}(b1), with the flow  frozen in the rotating frame forming a circle of equilibria in Fig. \ref{fig:NormalPoleFlip}(b2). Physically, this means that these periodic orbits conserve the semiclassical energy per atom $E = \braket{\hat{H}_{\mathrm{LMG}}}/N = \omega_0\gamma$. That is, in the (three-dimensional) LMG model description, after adiabatic elimination, the entire phase space $\mathbb{S}^{2}$ is foliated by periodic orbits, meaning that the conservation of energy is satisfied \emph{globally}. In contrast, in the full (four-dimensional) description of the Dicke model \cite{stitely_nonlinear_2020}, the periodic orbits foliate a two-dimensional (attracting) invariant surface within the full phase space $\mathbb{R}^2 \times \mathbb{S}^2$. As observed in Ref.~\cite{stitely_nonlinear_2020}, the conservation of energy at the pole-flip transition in the Dicke model is satisfied within the two-dimensional surface, but not globally. This implies that the two-dimensional surface observed in the Dicke model at the moment of the pole-flip transition is a signature of the semiclassical LMG model behavior discussed above.

\subsection{Effect of Atomic Dissipation and Pumping}\label{SubSect:Addition}

The addition of a pump that excites the atomic states removes the possibility for a normal phase with all spins down, and produces lasing behavior if the co-rotating coupling strength $\lambda_-$ is sufficiently large. Similarly, the inclusion of atomic dissipation through spontaneous emission removes the possibility of an inverted normal phase and can seed counter-lasing behavior with a strong counter-rotating coupling strength $\lambda_+$. We will now show that these lasing phases appear in the semiclassical model as stable periodic orbits arising from Hopf bifurcations. 

\subsubsection{Hopf Bifurcations to Counter-Lasing}

With nonzero atomic pumping $\Gamma_\uparrow$ and dissipation $\Gamma_\downarrow$ in the strongly dissipative model, angular momentum conservation (\ref{eqn:conservation}) is broken ($dr/dt \neq 0$); thus, freeing the dynamics from (the surface of) the Bloch sphere and removing the inverted normal phase equilibrium. Crucially, this means that system (\ref{eqn:ODEs1})--(\ref{eqn:ODEs2}) now has a \emph{single} equilibrium at
\begin{equation}\label{eqn:NormalPhaseEquilibrium}
\beta_\mathrm{eq} = 0,\ \gamma_{\mathrm{eq}} = \frac{1}{2}\frac{\Gamma_\uparrow - \Gamma_\downarrow}{\Gamma_\uparrow + \Gamma_\downarrow} = \frac{\delta}{2\sigma}.
\end{equation}
Note that if $\Gamma_\uparrow > \Gamma_\downarrow$, then $\gamma_{\mathrm{eq}}$ is strictly positive, and it is strictly negative if $\Gamma_\downarrow > \Gamma_\uparrow$.

\begin{figure}[t]
	\centering
	\includegraphics[width=9.6cm]{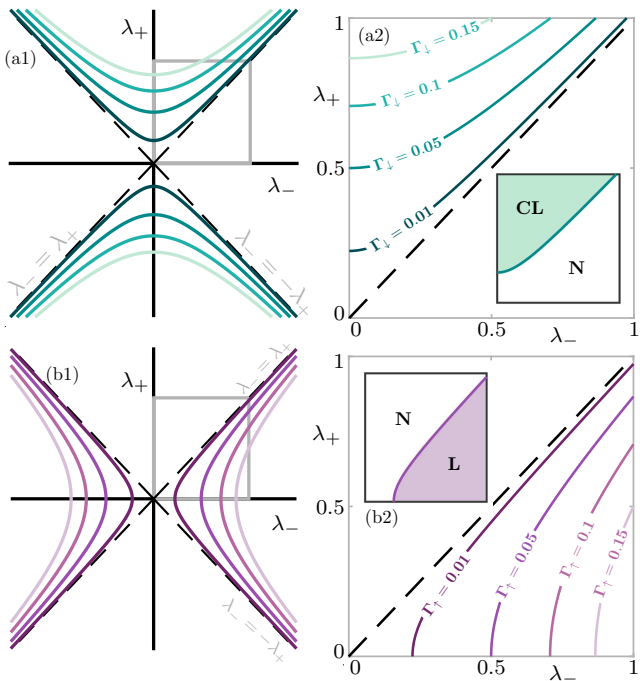}
	\caption{\label{fig:HopfHyperbola} Hopf bifurcation curves in the $(\lambda_{-},\lambda_+)$-plane. Panel (a) shows a selection of the hyperbolae for the counter-lasing transition when $\Gamma_\uparrow = 0$; here $\Gamma_\downarrow = 0.01,0.05,0.1,0.15$. Panel (b) shows the case of the lasing transition for $\Gamma_\downarrow = 0$ for $\Gamma_\uparrow = 0.01,0.05,0.1,0.15$. Panels (a1) and (b1) show the extension of the Hopf bifurcation curves for negative coupling strengths to highlight the symmetry properties of the curves. Panels (a2) and (b2) are enlargements and their insets show on which sides of the curves the counter-lasing/lasing and normal phases reside. Other parameters are $\kappa = 5$, $\omega_0 = 1$.}
\end{figure}

The Jacobian matrix $J$ of system (\ref{eqn:ODEs1})--(\ref{eqn:ODEs2}) in the strongly dissipative case at the equilibrium $\beta_\mathrm{eq} = 0$, $\gamma = \gamma_{\mathrm{eq}}$ has the simple (block diagonal) structure
\begin{equation}\label{eqn:jacobian}
	J = \begin{bmatrix}
		-\mu\gamma_{\mathrm{eq}} - \sigma & -\omega_0 & 0 \\
		\omega_0 & -\mu \gamma_{\mathrm{eq}}- \sigma & 0 \\
		0 & 0 & -2\sigma
	\end{bmatrix},
\end{equation}
with eigenvalues
\begin{align}
	v_{x,y} &= -\mu\gamma_{\mathrm{eq}} - \sigma\pm i\omega_0, \label{eqn:eval1} \\
	v_{\gamma} &= -2\sigma. \label{eqn:eval2}
\end{align}
The eigenvectors associated with the eigenvalues $v_{x,y}$ span the $\beta$-plane at a height $\gamma_\mathrm{eq}$. On the other hand, the eigenvector associated with $v_\gamma$ spans the $\gamma$-axis. Equation~(\ref{eqn:eval1}) for the eigenvalues $v_{x,y}$ indicates the presence of a Hopf bifurcation when $v_{x,y}$ form a purely imaginary pair $\pm i\omega_0$, that is, when
\begin{equation}\label{eqn:mu_hopf}
	\mu = \frac{-2\sigma^2}{\delta}.
\end{equation}
Or in terms of the LMG model parameters [Eqs. (\ref{eqn:parameters})],
\begin{equation}\label{eqn:HopfCurves}
\lambda_{+}^2 - \lambda_{-}^2 = \frac{-(\Gamma_\uparrow + \Gamma_\downarrow)^2}{\eta(\Gamma_\uparrow - \Gamma_\downarrow)},
\end{equation}
which describes a family of hyperbolae in the $(\lambda_{-},\lambda_{+})$-plane that asymptote onto the diagonal $\lambda_{-}=\lambda_{+}$. The Hopf bifurcations along these curves create circular periodic orbits, since system (\ref{eqn:SphereODEs1})--(\ref{eqn:SphereODEs3}) is $U(1)$-symmetric, that is, invariant under $\mathcal{S}$.

The periodic orbits formed from these Hopf bifurcations represent \emph{lasing states} (\textbf{L}) for $\Gamma_\uparrow > \Gamma_\downarrow$, and \emph{counter-lasing states} (\textbf{CL}) for $\Gamma_\uparrow < \Gamma_\downarrow$ \cite{kirton_superradiant_2018}. The Hopf bifurcation then marks the phase transition from the normal phase (\textbf{N}) to the lasing/counter-lasing phase. A selection of these Hopf bifurcation curves are shown in Fig.~\ref{fig:HopfHyperbola}(a) for the counter-lasing phase transition for $\Gamma_{\uparrow} = 0$ and $\Gamma_\downarrow = 0.01,0.05,0.1,0.15$, while Fig. \ref{fig:HopfHyperbola}(b) shows the bifurcation curves for the lasing transition for $\Gamma_\downarrow = 0$ and $\Gamma_\uparrow = 0.01,0.05,0.1,0.15$. These hyperbolae divide the $(\lambda_{-},\lambda_{+})$-plane into regions where the counter-lasing (\textbf{CL}), lasing (\textbf{L}), and normal (\textbf{N}) phases may be found. Importantly, the addition of atomic dissipation or pumping, $\Gamma_{\uparrow},\Gamma_\downarrow>0$, not only creates the possibility of lasing or counter-lasing but lifts the phase transition curve from the origin. This means that, for the case $\Gamma_\uparrow = 0$, there is now a nonzero minimum threshold value of $\lambda_{+}$ for the transition to occur. This was noticed experimentally in Ref. \cite{zhiqiang_nonequilibrium_2017} where, above a certain threshold of the ratio $\lambda_{+}/\lambda_{-}$ corresponding to the line of the pole-flip transition, no phase transition is expected if atomic damping is neglected. The minimum (so-called single-beam) threshold for the transition was also investigated experimentally in Ref. \cite{zhang_dicke-model_2018}. This minimum threshold occurs when either $\lambda_{-}$ or $\lambda_{+}$ is zero, depending on the sign of $\delta = \Gamma_\uparrow - \Gamma_\downarrow$. In the semiclassical LMG model, the minimum threshold occurs at
\begin{equation}
	\lambda_{\pm} = \frac{\Gamma_{\uparrow} + \Gamma_\downarrow}{\sqrt{\mp \eta(\Gamma_\uparrow - \Gamma_\downarrow)}}.
\end{equation}

We remark that in terms of Eqs. (\ref{eqn:SphereODEs1})--(\ref{eqn:SphereODEs2}) the Hopf bifurcation is manifested as a \emph{pitchfork bifurcation} of relative equilibria. This is easily seen since the normal phase equilibrium
\begin{equation}\label{eqn:relative}
(r,\theta) = 
\begin{cases}
(|\gamma_\mathrm{eq}|,0),\ \text{if } \delta > 0, \\
(|\gamma_\mathrm{eq}|,\pi),\ \text{if } \delta < 0, \\
(0,0),\ \text{if } \delta = 0,\ \sigma > 0,
\end{cases}
\end{equation}
always exists, and the relative equilibria, which represent periodic orbits, bifurcate from this equilibrium.
\begin{figure}[t]
	\centering
	\includegraphics[width=7.8cm]{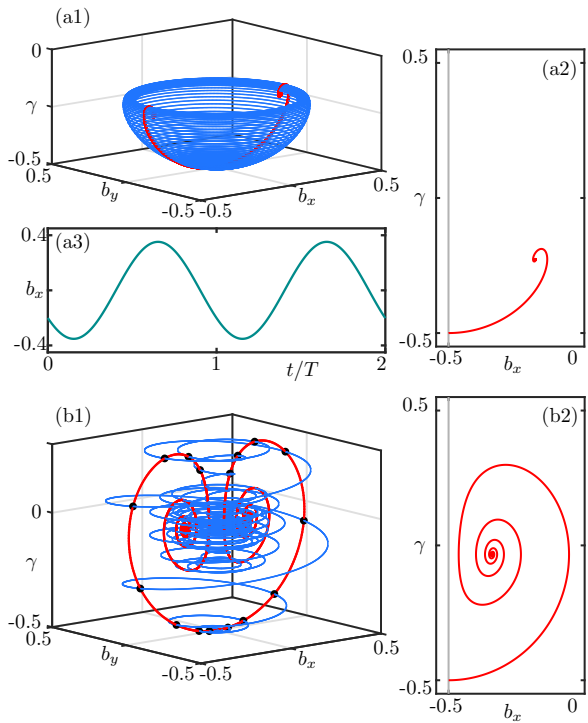}
	\caption{\label{fig:Converge2} Convergence of trajectories towards counter-lasing periodic orbits as $\lambda_{+}$ is varied, keeping $\Gamma_\downarrow = 0.01$ constant. Panel (a1) shows a trajectory initiated at $(r,\theta,\phi) = (1/2,\pi+0.001,0)$ in $(b_{x},b_{y},\gamma)$-space for $\omega_0 = 1$. Panel (a2) shows the trajectory in the rotating frame, where the periodic orbit appears as a relative equilibrium point. Panel (a3) shows the temporal trace of $b_{x}$ of the counter-lasing periodic orbit. Panel (b1) shows a trajectory approaching a counter-lasing periodic orbit further from the Hopf bifurcation with $\omega_0 = 1$, with the rotating frame trajectory shown in panel (b2). In panel (a) $\lambda_{+} = 0.6$ and in panel (b) $\lambda_{+} = 1$. For both, $\lambda_{-} = 0.5$, $\kappa = 5$, $\Gamma_\uparrow = 0$.}
\end{figure}
\begin{figure}[b]
	\centering
	\includegraphics[width=8.6cm]{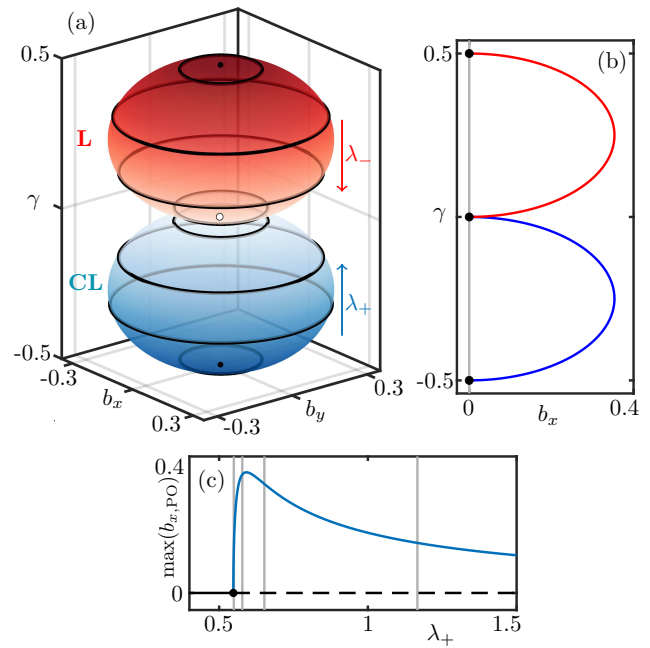}
	\caption{\label{fig:CL_orbits}Surfaces of bifurcating periodic orbits. Panel (a) shows two oblate spheroids of periodic orbits formed after the Hopf bifurcation. On the top in red is the spheroid formed from the lasing orbits for $\lambda_{+} = 0.5$, $\Gamma_\uparrow = 0.01$, $\Gamma_\downarrow = 0$, with increasing $\lambda_{-}$ down the surface towards the origin, shown as a white dot. On the bottom in blue is the spheroid formed from counter-lasing periodic orbits for $\lambda_{-} = 0.5$, $\Gamma_\uparrow = 0$, $\Gamma_\downarrow = 0.01$, with increasing $\lambda_{+}$ up the surface towards the origin. Panel (b) shows the projections of these surfaces of periodic orbits on the half-plane $(b_{x},\gamma)$. Panel (c) shows $\max(b_{x,\mathrm{PO}})$ of the periodic orbits for the counter-lasing orbits as $\lambda_{+}$ is increased. The dashed line indicates where the normal phase equilibrium point is unstable; vertical gray lines indicate the periodic orbits highlighted on the spheroids in panel (a). The curve of $\max(b_{x,\mathrm{PO}})$ for the lasing orbits is the same as for the counter-lasing orbits, but for increasing $\lambda_{-}$ rather than $\lambda_{+}$. Other parameters are $\kappa = 5$, $\omega = 0$, $\omega_0 = 1$.}
\end{figure}

\subsubsection{Transient Dynamics}

The system's convergence to a counter-lasing periodic orbit formed just after the Hopf bifurcation is shown in Fig. \ref{fig:Converge2}(a1) by numerical integration of Eqs. (\ref{eqn:SphereODEs1})--(\ref{eqn:SphereODEs3}), with initial conditions perturbed from the (linearly unstable) equilibrium point $(r,\theta) = (0.5,\pi)$. Due to the constant rotation by the frequency $\omega_0$ about the $\gamma$-axis, the ``density" of intersection points on an (arbitrary) plane defined by $b_y = 0$ gives information about the speed of convergence of the trajectory onto the periodic orbit, since the trajectory will intersect the plane every $T = 2\pi/\omega_0$ units of time. Thus, it can be seen that, just after the Hopf bifurcation, the spiralling trajectory in Fig. \ref{fig:Converge2}(a1) approaches the periodic orbit slowly. The corresponding trajectory in the rotating frame is shown in Fig. \ref{fig:Converge2}(a2), highlighting the tightly-wound spiral structure of the flow, while Fig. \ref{fig:Converge2}(a3) is a time series of the attracting periodic orbit. Further away from the Hopf bifurcation, as $\lambda_{+}$ is increased, convergence to the periodic orbit is initially faster in Fig. \ref{fig:Converge2}(b1) and the spiralling structure in Fig. \ref{fig:Converge2}(b2) is more pronounced.

\subsubsection{Geometry of Counter-Lasing Orbits}

The geometry of the periodic orbits describing the lasing and counter-lasing phases can be examined with Eqs.~(\ref{eqn:SphereODEs1})--(\ref{eqn:SphereODEs3}) as $\mu$ is varied. As mentioned previously, the invariance under the transformation $\mathcal{S}$ ensures the periodic orbits are circular and have period $T=2\pi/\omega_0$. A periodic orbit with radius $r_{\mathrm{PO}}$ and polar angle $\theta_{\mathrm{PO}}$ must have $dr/dt = d\theta/dt = 0$, that is, periodic orbits are described by relative equilibria of the equations in the rotating frame, Eqs. (\ref{eqn:SphereODEs1})--(\ref{eqn:SphereODEs2}). Solving for them as common roots gives
\begin{equation}\label{eqn:cos_theta}
	\cos(\theta_{\mathrm{PO}}) = \frac{\sigma}{\sqrt{-(\delta\mu + \sigma^2)}}.
\end{equation}
This equation sets the condition for relative equilibria in addition to Eq.~(\ref{eqn:relative}) to exist, since $-1 \leq \cos(\theta_\mathrm{PO}) \leq 1$. Either equality is attained, for $\delta>0$ or $\delta<0$, respectively, when Eq.~(\ref{eqn:mu_hopf}) is satisfied. This describes a pitchfork bifurcation of Eqs. (\ref{eqn:SphereODEs1})--(\ref{eqn:SphereODEs2}) as the relative equilibria $(r_\mathrm{PO},\theta_\mathrm{PO})$ emerge from the normal phase equilibrium with an exchange of stability. The radius $r_{\mathrm{PO}}$ is
\begin{equation}\label{eqn:r}
r_\mathrm{PO} = \frac{\delta\cos(\theta_\mathrm{PO})}{\sigma\left[ 1 + \cos^{2}(\theta_\mathrm{PO}) \right]} = \frac{\sqrt{-(\delta\mu + \sigma^2)}}{\mu},
\end{equation}
which also illustrates the Hopf bifurcation condition Eq.~(\ref{eqn:mu_hopf}) when $r_\mathrm{PO} = 0$.

\begin{figure}[t]
	\centering
	\includegraphics[width=8cm]{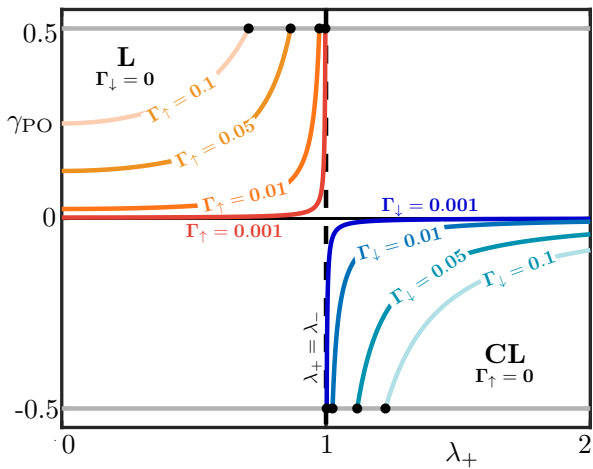}
	\caption{\label{fig:BifConverge} Bifurcation diagram of periodic orbits formed from Hopf bifurcations as $\lambda_{+}$ is increased. The blue colored curves on the right are curves in the counter-lasing phase: $\Gamma_\downarrow = 0.001,0.01,0.05,0.1$, $\Gamma_\uparrow=0$. The orange colored curves on the left are for the lasing phase: $\Gamma_\downarrow = 0$, $\Gamma_\uparrow = 0.001,0.01,0.05,0.1$. Black dots indicate Hopf bifurcations. Other parameters are $\kappa = 5$, $\lambda_{-} = 1$.}
\end{figure}

The first equality in Eq.~(\ref{eqn:r}) describes an \emph{oblate spheroid} centered at $(b_x,b_y,\gamma)=(0,0,\gamma_\mathrm{eq}/2)$, an ellipsoid with equal axes in planes of constant $\gamma$, parameterized by $\mu$. Two surfaces of periodic orbits for the lasing (\textbf{L}) and counter-lasing (\textbf{CL}) cases are shown in Fig. \ref{fig:CL_orbits}(a), for constant $\lambda_{+}$ and varying $\lambda_-$ (\textbf{L} surface) and for constant $\lambda_-$ and varying $\lambda_+$ (\textbf{CL} surface), the projection of the surface onto the half-plane $(b_{x},\gamma)$ is shown in Fig.~ \ref{fig:CL_orbits}(b). For the surface \textbf{L} of lasing orbits, the Hopf bifurcation occurs at the top where $\gamma = \gamma_{\mathrm{eq}}>0$; as $\lambda_{-}$ is increased the orbits move down the surface towards the origin as $\lambda_{-}\rightarrow\infty$. For the counter-lasing orbits, the Hopf bifurcation occurs at the bottom of the surface \textbf{CL} where $\gamma = \gamma_\mathrm{eq} < 0$; as $\lambda_{+}$ is increased the orbits move up the surface, again towards the origin as $\lambda_{+}\rightarrow\infty$. The semi-minor and semi-major axes of the spheroid are
\begin{equation}
	a = \frac{\delta}{4\sigma}\ \text{and} \ b = \frac{\delta}{2\sqrt{2}\sigma},
\end{equation}
respectively. The poles along the $\gamma$-axis are $\gamma = \gamma_\mathrm{eq}$, where the Hopf bifurcation occurs, and $\gamma=0$, where the periodic orbits accumulate as $\mu\rightarrow\infty$. We remark that even though the adiabatic approximation in the large coupling limit is not strictly satisfied, the consideration of such limits give insight into how the system behaves when the coherent processes strongly dominate.

From Eqs. (\ref{eqn:cos_theta})--(\ref{eqn:r}), the family of periodic orbits emerging from the Hopf bifurcation can be described analytically as $\mu$ varies. The periodic orbits are characterized by the maxima (or minima) of either $b_x$ or $b_y$ (we only need one due to $U(1)$ symmetry) and $\gamma$. Firstly, we find
\begin{equation}
	\max(b_{x,\mathrm{PO}}) = r_\mathrm{PO}\sin(\theta_\mathrm{PO}) = \frac{\sqrt{2\eta\delta(\lambda_{-}^2 - \lambda_{+}^2) - 2\sigma^2}}{2\eta(\lambda_{+}^2 - \lambda_{-}^2)},
\end{equation}
which is illustrated, as $\lambda_{+}$ is varied, in Fig. \ref{fig:CL_orbits}(c) for the counter-lasing transition. Here, we see a very fast initial growth with $\lambda_+$ in the size of the periodic orbits (as a function of $\lambda_{+}$) after the Hopf bifurcation, followed by a slow relaxation as the periodic orbits shrink towards the origin as $\lambda_{+}\rightarrow\infty$. This indicates that, once the Hopf bifurcation is reached, the periodic orbits move along the surface of periodic orbits in Fig. \ref{fig:CL_orbits}(a) very quickly upon variation of $\lambda_{+}$, before slowing down as they approach the origin, at the top of the surface where $\gamma = 0$. This applies to the lasing transition as well, where $\Gamma_\uparrow > \Gamma_\downarrow$ and the curves are now functions of $\lambda_{-}$, with $\lambda_{+}$ held constant.

\subsection{Atomic Dissipation and Pumping near the Pole-Flip Limit}

\begin{figure}[t]
	\centering
	\includegraphics[width=7.8cm]{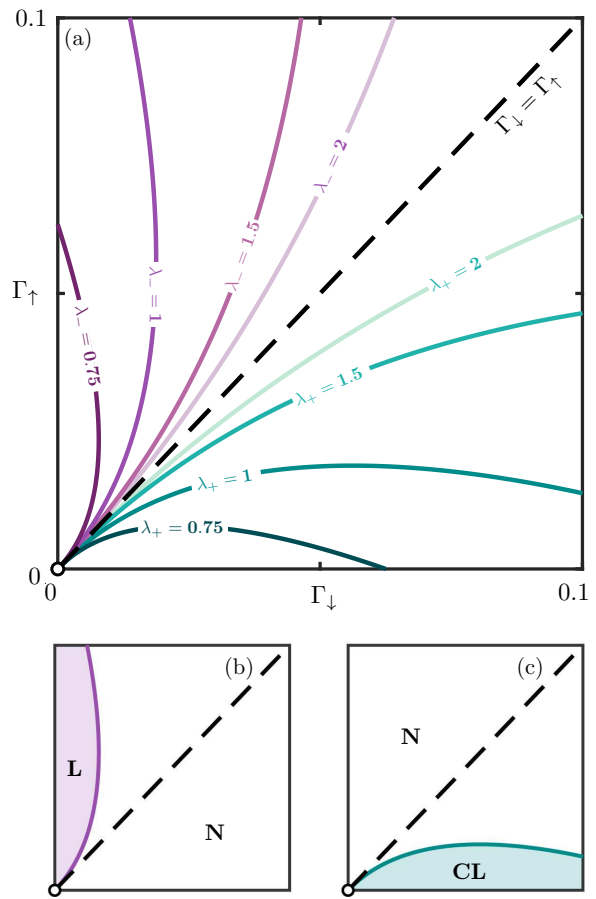}
	\caption{\label{fig:GammaPlaneHopf}Hopf bifurcation curves in the $(\Gamma_\downarrow,\Gamma_\uparrow)$-plane. In panel (a), the green colored curves on the right are the Hopf bifurcation curves that lead to the creation of counter-lasing periodic orbits for $\lambda_{-} = 0.5$, $\lambda_{+} = 0.75, 1,1.5,2$, while the purple colored curves on the left lead to the creation of lasing periodic orbits for $\lambda_{+} = 0.5, \lambda_{-} = 0.75, 1, 1.5, 2$. Panels (b) and (c) show how these curves separate the $(\Gamma_\downarrow,\Gamma_\uparrow)$-plane into regions where the normal (\textbf{N}), lasing (\textbf{L}), and counter-lasing (\textbf{CL}) phases reside. All of the curves are for $\kappa = 5$.}
\end{figure}

Information about how the system approaches the pole-flip transition in the limit $\delta,\sigma\rightarrow 0$ can be obtained from the $\gamma$-components of the periodic orbits,
\begin{equation}
\gamma_{\mathrm{PO}} = r_\mathrm{PO}\cos(\theta_\mathrm{PO}) = \frac{\sigma}{\mu} = \frac{\Gamma_\uparrow + \Gamma_\downarrow}{2\eta(\lambda_{+}^2 - \lambda_{-}^2)},
\end{equation}
which are shown in Fig. \ref{fig:BifConverge} as $\lambda_{+}$ is varied in the lasing phase for $\Gamma_\downarrow = 0$, $\Gamma_\uparrow = 0.001,0.01,0.05,0.1$, and in the counter-lasing phase for $\Gamma_\uparrow = 0$, $\Gamma_\downarrow = 0.001,0.01,0.05,0.1$. The curves, which are hyperbolae, in the upper left hand side correspond to the periodic orbits in the lasing phase, where $\Gamma_\uparrow>\Gamma_\downarrow$ and $\lambda_{-}>\lambda_{+}$. As $\lambda_{+}$ is increased these orbits shrink and disappear at Hopf bifurcations, causing the disappearance of the lasing phase and the appearance of the normal phase. On the other hand, the curves on the bottom right-hand side correspond to periodic orbits in the counter-lasing phase, where $\Gamma_\downarrow > \Gamma_\uparrow$ and $\lambda_{+} > \lambda_{-}$. In this case, upon increase of $\lambda_{+}$ the normal phase is destabilized and the counter-lasing phase emerges. In the limit $\Gamma_\downarrow, \Gamma_\uparrow \rightarrow 0$ all of these curves accumulate onto the vertical dashed line at $\lambda_{-} = \lambda_{+}$, which corresponds to the pole-flip transition, and indicates coexistent periodic orbits for all values of $\gamma$ between -0.5 and 0.5 [see Fig. \ref{fig:NormalPoleFlip}(b)].

The two-parameter bifurcation diagram in the $(\Gamma_\downarrow,\Gamma_\uparrow)$-plane can be obtained from Eq. (\ref{eqn:HopfCurves}), solved with the coupling strengths $\lambda_{-}$ and $\lambda_{+}$ kept constant with $\Gamma_\downarrow$ and $\Gamma_\uparrow$ now allowed to vary. Here the Hopf bifurcation curves form a family of parabolae in the $(\Gamma_\downarrow,\Gamma_\uparrow)$-plane, which intersect the origin tangentially to the diagonal $\Gamma_\downarrow = \Gamma_\uparrow$ and intersect $\Gamma_\uparrow=0$ at $\Gamma_\downarrow = \mu$ in the counter-lasing phase; and similarly intersect $\Gamma_\downarrow = 0$ at $\Gamma_\uparrow = -\mu$ in the lasing phase. A selection of these bifurcation curves for various values of the coupling strengths are shown in Fig. \ref{fig:GammaPlaneHopf}(a). The curves in the bottom right portion of the diagram are the Hopf bifurcation curves that give rise to the periodic orbits in the counter-lasing phase, likewise the curves in the top left portion create the lasing phase periodic orbits. Along the (positive) diagonal $\Gamma_\downarrow = \Gamma_\uparrow > 0$ the normal phase equilibrium is located at the origin and is always stable. The counter-lasing Hopf bifurcation curves accumulate onto the diagonal in the limit $\lambda_{+}\rightarrow\infty$, as do the lasing Hopf bifurcation curves for $\lambda_{-}\rightarrow\infty$.

\begin{figure}[t]
	\centering
	\includegraphics[width=8.3cm]{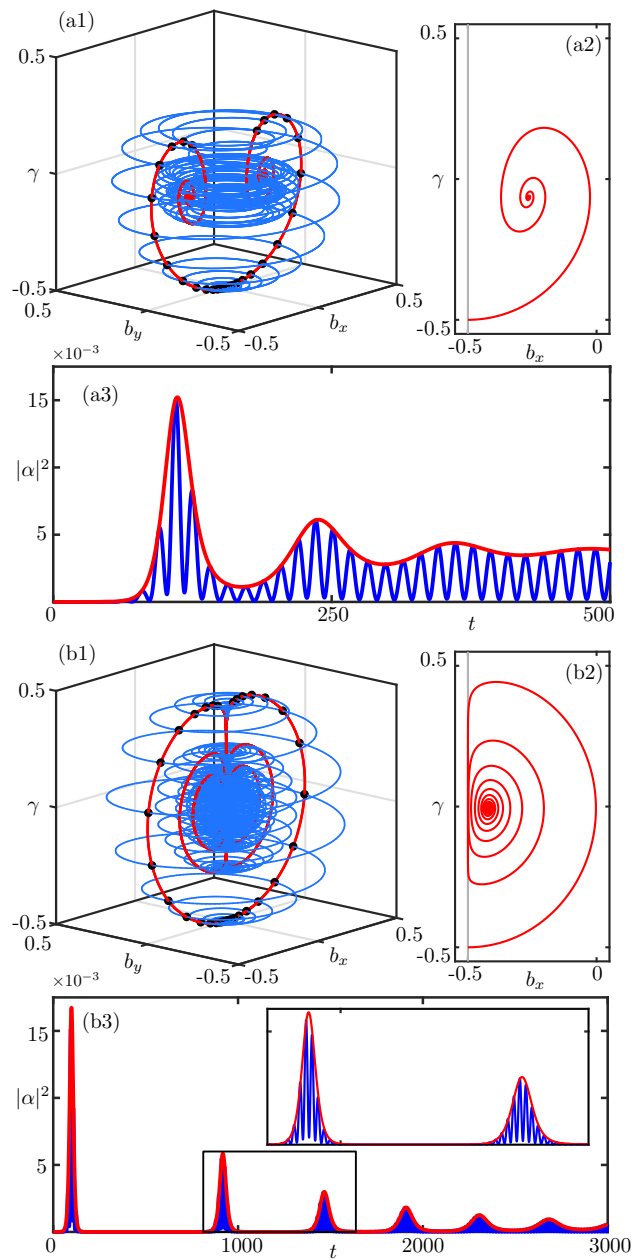}
	\caption{\label{fig:Converge1} Transient dynamics showing how the counter-lasing periodic orbits are approached. Panels (a1) and (b1) show a simulation of Eqs. (\ref{eqn:SphereODEs1})--(\ref{eqn:SphereODEs3}) in atomic space $(b_x,b_y,\gamma)$ for $\omega_0 = 1$. In panels (a) $\Gamma_\downarrow = 0.01$, in panels (b) $\Gamma_\downarrow = 0.001$. Panels (a2) and (b2) show the trajectories in the rotating frame that describe the convergence to the periodic orbit, corresponding to (a1) and (b1), respectively. In panels (a3) and (b3) we show the temporal trace of the photon number $|\alpha|^2$ for $\omega_0 = 0.2$. For both $\lambda_{-} = 0.5$, $\lambda_{+} = 0.8, \kappa = 5$, $\Gamma_\uparrow = 0$.}
\end{figure}

The Hopf bifurcation curves approach the origin $\Gamma_\downarrow = \Gamma_\uparrow = 0$. At this point, the situation described in Sect.~\ref{SubSect:NormalPoleFlip} is regained when angular momentum is conserved and the $\gamma$-axis is a one-dimensional manifold of equilibria. The dynamics here are highly degenerate. According to Eqs. (\ref{eqn:eval1}) and (\ref{eqn:eval2}), the eigenvalues of these equilibria are $v_{x,y} = -\mu\gamma\pm i\omega_0$ and $v_\gamma = 0$, indicating a zero eigenvalue for all $\mu$. There are two configurations this line of equilibria can take, depending on the sign of $\mu$. For $\mu < 0$, the equilibria with $\gamma < 0$ are stable (in the $(b_{x},b_{y})$-plane), while for $\gamma > 0$ they are unstable. Here the system is in the spin-down normal phase. The change in stability that takes place at the origin $b_{x} = b_{y} = \gamma = 0$ is a bifurcation parameterized not by a system parameter, but by the \emph{coordinate} $\gamma$. This is an example of a \emph{bifurcation without parameters} \cite{liebscher_bifurcation_2015}. If $\mu > 0$ the opposite configuration occurs: equilibria with $\gamma < 0$ are unstable, and those with $\gamma > 0$ are stable. Here the system is in the spin-up normal phase. The transition between these two configurations occurs for $\mu = 0$, when each equilibrium features the eigenvalues $v_{x,y} = \pm i\omega_0$ and a zero eigenvalue $v_{\gamma} = 0$. Unlike the generic case of the zero-Hopf bifurcation \cite{kuznetsov_elements_2004}, which satisfies this eigenvalue configuration, the particular situation we are considering leads to the creation of infinitely many coexisting periodic orbits that foliate phase space, as we saw before in Fig.~\ref{fig:NormalPoleFlip}(b). This means that the case we are observing in the semiclasscal LMG model corresponds to an unfolding of a degenerate zero-Hopf bifurcation.

The bifurcation analysis of the emergence of the periodic orbits corresponding to the lasing and counter-lasing phases above considers only the final states as $t\rightarrow\infty$. However, in many situations (barring the addition of laser fields tuned to transitions between the atomic ground states) the spontaneous emission rates $\Gamma_\downarrow$ are much smaller than the rates that determine the interactions $\lambda_{\pm}$ or the cavity decay rate $\kappa$. As a result, there are two relevant timescales when the atomic damping and pumping is small.

To examine how trajectories converge onto the periodic orbits in the counter-lasing phase, we numerically integrate Eqs. (\ref{eqn:SphereODEs1})--(\ref{eqn:SphereODEs3}) with an initial condition perturbed from the (linearly unstable) normal phase equilibrium $(r,\theta) = (0.5,\pi)$. The results are plotted in Fig.~\ref{fig:Converge1}. In panel (a1) we show the transient convergence to the counter-lasing periodic orbit for $\Gamma_\downarrow = 0.01$. The trajectory spirals radially inwards to the periodic orbit, accompanied by angular rotation at frequency $\omega_0$. Again, the spiralling motion is more clearly seen in the rotating frame in panel (a2). This spiralling behavior of the atomic dynamics also appears in the cavity field dynamics, as the cavity field amplitude $\alpha=\braket{\hat{a}}/\sqrt{N}$ is directly related to $\beta$. The photon number is
\begin{equation}
	|\alpha|^2 = (\xi^2 + \eta^2)\left[ (\lambda_{-}^2 + \lambda_{+}^2)\beta^*\beta + \lambda_{-}\lambda_{+}(\beta^2 + {\beta^*}^2) \right],
\end{equation}
which is shown in Fig. \ref{fig:Converge1}(a3) (with $\omega_0=0.2$ for clarity of illustration). The inwards spiralling of the rotating frame trajectories modulates the amplitude of oscillations of the photon number, which eventually settles to a periodic oscillation in the counter-lasing phase. The photon number in the rotating frame provides the amplitude envelope.  

Figure \ref{fig:Converge1}(b) shows the full and rotating frame trajectories as well as the temporal trace of the photon number when the rate of spontaneous emission is even smaller at $\Gamma_\downarrow = 0.001$. In Fig. \ref{fig:Converge1}(b1) the trajectory is initially perturbed from the normal phase equilibrium and makes a rapid transition closely following along the Bloch sphere of radius $1/2$, to its North pole. Here the dynamics slow down dramatically and the trajectory slowly descends along the $\gamma$-axis; this axis acts as a slow manifold, since the timescale of motion in this region of phase space is given only by $\Gamma_\downarrow$ and $\Gamma_\uparrow$. After the slow descent along the $\gamma$-axis, the trajectory lifts off and makes another rapid traversal along another Bloch sphere ``remnant." Again, the geometry of how the periodic orbit is approached is more clear in the rotating frame trajectory illustrated in Fig. \ref{fig:Converge1}(b2). There is a sequence of rapid movement up Bloch sphere remnants as the periodic orbit is approached, each followed by slow descents along the $\gamma$-axis. The contrast between the slow movements along the $\gamma$-axis and the fast movements away is clearly demonstrated in the photon number, shown in Fig. \ref{fig:Converge1}(b3). Each rapid transition along a Bloch sphere remnant is associated with a pulse of photons, followed by a relaxation period. As a result, the lift-off of the trajectory from the $\gamma$-axis produces resurgent superradiant pulses. Each subsequent pulse has a larger width and a smaller amplitude as the counter-lasing phase is approached. 

In the parameter regime we are considering here, the light field in the full Dicke model description adiabatically follows the motion of the spins, such that each rapid traversal along each shell produces a pulse of photons as the spins are excited. The first traversal reaches the highest level of excitation, and so produces the highest amplitude light pulse. Moreover, each pulse arrives sooner than the last, until there are no identifiable pulses left when the periodic orbit is reached. It is the first of these photon pulses that are observed experimentally in Ref. \cite{zhiqiang_nonequilibrium_2017} to determine when the system is in the inverted phase. Experiments typically occur on a timescale such that only one pulse can be observed, due to atom loss or dephasing. However, when the counter-lasing periodic orbit is large (closer to the Hopf bifurcation), trajectories are not compressed along the slow manifold [see Fig.~\ref{fig:Converge2}(a2)], so that the separation of timescales that leads to the creation of clearly separated photon pulses is absent. This leads to a lengthening in time of the photon pulses and a sharp peak in the Fourier spectrum near the phase transition threshold. A similar effect has been noticed experimentally in Ref. \cite{zhiqiang_nonequilibrium_2017}.

%
%

\section{Transition between counter-lasing and superradiance}

We now consider the full Eqs. (\ref{eqn:ODEs1})--(\ref{eqn:ODEs2}) for $\xi \neq 0$, where the system loses its $U(1)$-symmetry and is only $\mathbb{Z}_{2}$ invariant. Hence, there is now the possibility to spontaneously break the $\mathbb{Z}_{2}$ symmetry in a superradiant phase transition upon variation of $\lambda_+$. In this Section, we will show that for fixed coupling $\lambda_{-}$ and increasing $\lambda_{+}$, the counter-lasing phase must eventually stabilize even in the presence of superradiance. It is then imperative to understand how the counter-lasing phase emerges out of superradiance.

\subsection{Emergence of Superradiance}

The emergence of the superradiant phase corresponds to the appearance of stable asymmetric equilibria, which we will refer to as \emph{superradiant equilibria}. Much like its Dicke model counterpart, the transition to superradiance in the LMG model with unbalanced coupling can occur in two ways when varying $\lambda_{+}$, depending on the values of the other parameters (see Ref. \cite{stitely_nonlinear_2020} for details): 
\begin{enumerate}[leftmargin=2cm,itemsep=-4mm]
	\item[Case (i):] A supercritical pitchfork bifurcation $\mathrm{P}_\mathrm{spr}$ leads to the creation of two stable superradiant equilibria, whilst simultaneously turning the normal phase equilibrium unstable. \\
	\item[Case (ii):] A saddle-node bifurcation SN leads to the creation of two pairs of superradiant equilibria, two of which are stable and two of which are unstable. Subsequently, the two unstable superradiant equilibria collide with the stable normal phase equilibrium in a subcritical pitchfork bifurcation $\mathrm{P}_\mathrm{sub}$, where the unstable superradiant equilibria disappear and the normal phase becomes unstable; this leaves the other two superradiant equilibria as the only remaining stable objects. 
\end{enumerate}
Case (ii) leads to a parameter range of coexisting superradiant and normal phases, in between the saddle-node and pitchfork bifurcations, with an associated hysteresis loop and switching dynamics between the phases in the presence of quantum fluctuations \cite{stitely_superradiant_2020}. Such a hysteresis loop between the coexistent normal and superradiant phases has recently been observed experimentally in Ref. \cite{ferri_emerging_2021}.
\subsection{Superradiant Pole-Flip Transition}

\begin{figure}[t]
	\centering
	\includegraphics[width=8.6cm]{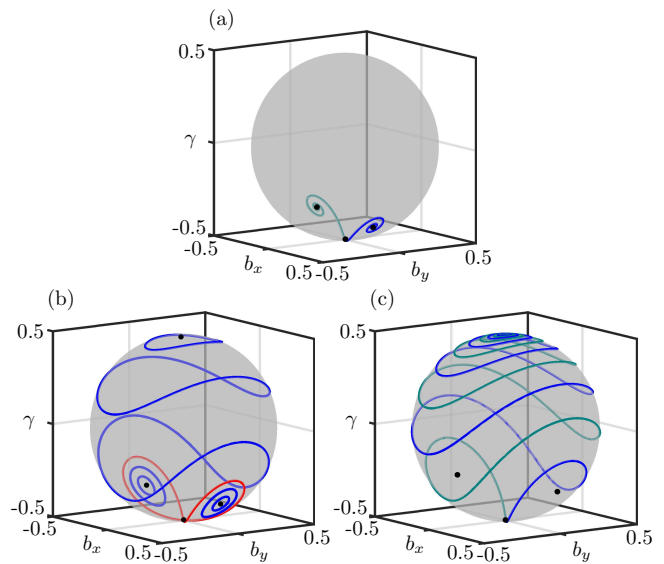}
	\caption{\label{fig:SRPoleFlip}Superradiant pole-flip transition on the Bloch sphere in the generalized semiclassical LMG model. In panel (a) the system is in the superradiant phase for $\lambda_{+} = 1.45$. During the superradiant pole-flip transition in panel (b), periodic orbits oscillating around the spin-up normal phase equilibrium and superradiant equilibria foliate the Bloch sphere, separated by a homoclinic orbit (shown in red) forming a figure eight through the spin-down normal phase equilibrium for $\lambda_{+} = 1.5$. In panel (c) the system is in the spin-up normal phase for $\lambda_{+} = 1.55$. In panels (a) and (c) trajectories are initialized along the unstable eigenspace of the normal phase equilibrium point. Other parameters are $\lambda_{-} = 1.5$, $\kappa = 4$, $\omega = 0.5$, $\omega_0 = 0.2$, $\Gamma_\uparrow = \Gamma_\downarrow = 0$.}
\end{figure}

In the absence of atomic dissipation and pumping, there is a similar transition to that as described in Sec. \ref{SubSect:NormalPoleFlip} where the inverted normal phase arises from the superradiant phase via a foliation of periodic orbits on the Bloch sphere \cite{stitely_nonlinear_2020}. This transition, called the \emph{superradiant pole-flip transition}, is illustrated in Fig. \ref{fig:SRPoleFlip}. In Fig.~\ref{fig:SRPoleFlip}(a), the system begins in the superradiant phase, where the asymmetric equilibria are stable. At the superradiant pole-flip transition point, the Bloch sphere is again foliated by periodic orbits, but with a different foliation: there are now periodic orbits established around each superradiant equilibrium point, as well as a pair of \emph{homoclinic orbits}, trajectories that converges to the (saddle-type) normal phase equilibrium forwards and backwards in time; see Fig. \ref{fig:SRPoleFlip}(b). As $\lambda_{+}$ is increased, the superradiant equilibria become unstable and the spin-up normal phase equilibrium (North pole of the Bloch sphere) becomes stable; see Fig. \ref{fig:SRPoleFlip}(c).

\subsection{Inclusion of Atomic Dissipation}

\begin{figure}[h]
	\centering
	\includegraphics[width=8cm]{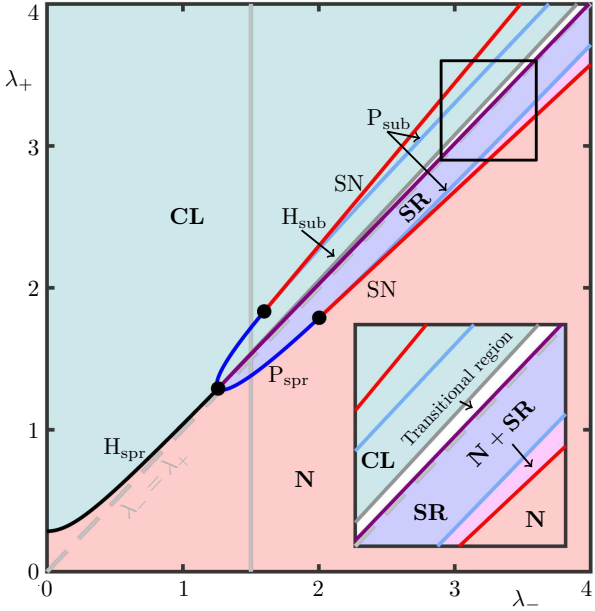}
	\caption{\label{fig:CouplingPlane} Two-parameter phase diagram in the ($\lambda_-$,$\lambda_+$)-plane for $\kappa = 4$, $\omega = 0.5$, $\omega_0 = 0.2$, $\Gamma_{\uparrow} = 0$, $\Gamma_\downarrow = 0.02$. Shown are curves of Hopf bifurcations $\mathrm{H}_{\mathrm{sub}}$ and $\mathrm{H}_{\mathrm{spr}}$ (gray and black curves), pitchfork bifurcations $\mathrm{P}_\mathrm{sub}$ and $\mathrm{P}_\mathrm{spr}$ (light and dark blue curves), and saddle-node bifurcations SN (red curves). Coloring shows the stable regions of the normal phase \textbf{N} (red region), the superradiant phase \textbf{SR} (blue region), coexistent normal and superradiant phases $\mathbf{N}+\mathbf{SR}$ (magenta region), and the counter-lasing phase \textbf{CL} (light-green region).}
\end{figure}

\begin{figure}[t]
	\centering
	\includegraphics[width=7.46cm]{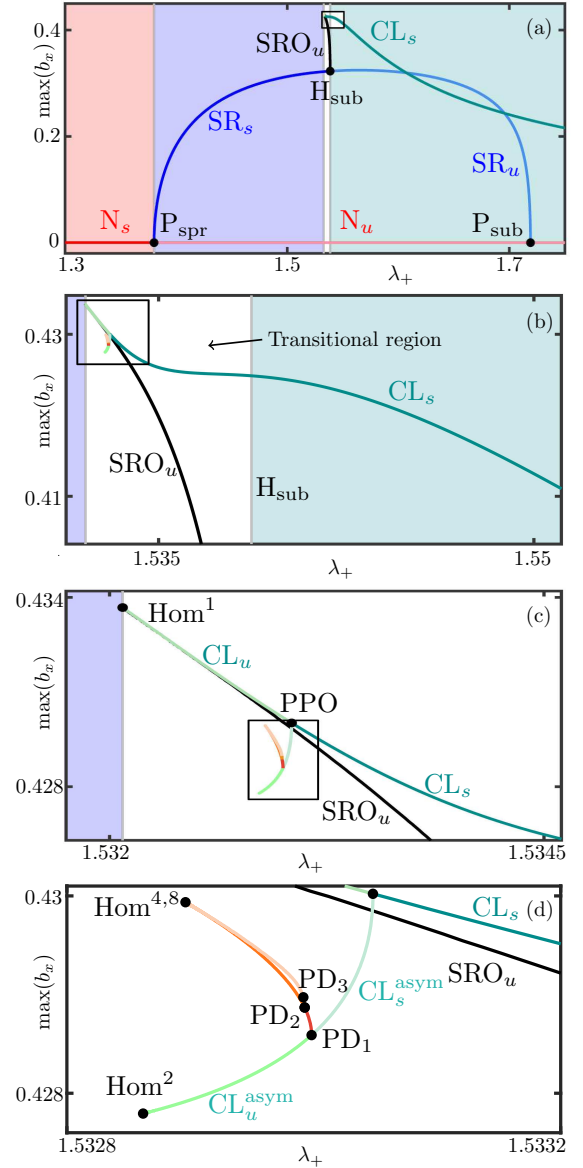}
	\caption{\label{fig:OneParBifDiag} One-parameter bifurcation diagram illustrating phase transitions from the normal phase, through the superradiant phase, into the counter-lasing phase for $\kappa = 4$, $\omega = 0.5$, $\omega_0 = 0.2$, $\Gamma_{\uparrow} = 0$, $\Gamma_\downarrow = 0.02$, $\lambda_{-} = 1.5$. Branches with a change in stability are unstable on the lighter-colored portion of the branch. The bifurcation diagram is divided up into colored regions where the system is in the normal phase (red), the superradiant phase (blue), a transitional region (white) between superradiance and counter-lasing, and where the system is entirely counter-lasing (light green). Each subsequent panel is a magnification of the rectangular region of the previous panel.}
\end{figure}

As in Sec. \ref{SubSect:Addition}, the addition of nonzero $\Gamma_\downarrow$ destroys the spin-up normal phase equilibrium and breaks the conservation law Eq. (\ref{eqn:conservation}). However, when $\xi \neq 0$ the situation is further complicated by the possibility of breaking the $\mathbb{Z}_2$ symmetry, leading to superradiance. If $\lambda_{-}$ is small, then varying $\lambda_{+}$ can lead to a transition from the normal phase to the counter-lasing phase as in Sect.~\ref{SubSect:NormalPoleFlip}, that is, via a Hopf bifurcation of the normal phase equilibrium point. However, the counter-lasing periodic orbits are no longer $U(1)$-symmetric.

\subsubsection{Phase Diagram in the Coupling Plane}

The addition of nonzero $\Gamma_\downarrow$ does not qualitatively change either case of the superradiant phase transition; it only shifts the phase transition boundaries. The two cases of the superradiant phase transition, as well as the Hopf bifurcation leading to the counter-lasing phase for small $\lambda_{-}$ can be summarized in one figure by computing the curves of pitchfork, saddle-node, and Hopf bifurcations in the $(\lambda_{-},\lambda_{+})$-plane. The details of the calculations of these curves are given in Appendix A. The resulting bifurcation or phase diagram is illustrated in Fig.~\ref{fig:CouplingPlane}. We note here that for the case $\Gamma_\downarrow = 0$ the pitchfork and saddle-node bifurcation curves are symmetric about the diagonal $\lambda_{-}=\lambda_{+}$, but the addition of $\Gamma_\downarrow\neq 0$ breaks this parameter symmetry also. Here we choose the parameter values $\kappa = 4$, $\omega = 0.5$, $\omega_0 = 0.2$, $\Gamma_\downarrow = 0.02$, and $\Gamma_\uparrow = 0$ so that the $\lambda_{+}$ values of the bifurcations we are concerned with are not too large. Otherwise, the adiabatic elimination procedure used to derive the LMG Hamiltonian Eq.~(\ref{eqn:HamiltonianLMG}) from the Dicke Hamiltonian Eq.~(\ref{eqn:DickeHamiltonian}) is no longer valid. The phase diagram shows the super- and subcritical Hopf bifurcation curves, $\mathrm{H}_{\mathrm{spr}}$ and $\mathrm{H}_{\mathrm{sub}}$, the super- and subcritical pitchfork bifurcation curves, $\mathrm{P}_\mathrm{spr}$ and $\mathrm{P}_\mathrm{sub}$, and a pair of saddle-node bifurcation curves SN. Also shown in Fig.~\ref{fig:OneParBifDiag} is an additional ``transitional region," which we will discuss later in the section. These bifurcation curves divide the phase diagram into regions where the normal \textbf{N}, superradiant \textbf{SR}, and counter-lasing \textbf{CL} phases are stable. 

Figure \ref{fig:CouplingPlane} outlines several mechanisms by which the counter-lasing phase \textbf{CL} is arrived at via a sequence of phase transitions from the normal phase \textbf{N} when $\lambda_{-}$ is kept constant and $\lambda_{+}$ is varied. When $\lambda_{-}$ is small, the transition from the normal phase to the counter-lasing phase occurs via a supercritical Hopf bifurcation $\mathrm{H}_\mathrm{spr}$. This is the same transition as in Sect. \ref{SubSect:Addition}, but now with the $U(1)$ symmetry explicitly broken. If $\lambda_{-}$ is larger, then upon increasing $\lambda_{+}$ the $\mathbb{Z}_{2}$-symmetry is spontaneously broken in a supercritical pitchfork bifurcation $\mathrm{P}_\mathrm{spr}$ after which the system is in the superradiant phase \textbf{SR}. This is case (i) of the superradiant phase transition. For even larger $\lambda_{-}$, we encounter case (ii). Here the superradiant phase emerges via a pair of saddle-node bifurcations SN and the system becomes multistable, with simultaneously stable normal and superradiant phases, in the region $\textbf{N}+\textbf{SR}$ shown in the inset panel of Fig.~\ref{fig:CouplingPlane}. The normal phase is then destabilized at the subcritical pitchfork bifurcation $\mathrm{P}_\mathrm{sub}$, after which the system is entirely in the superradiant phase \textbf{SR}. Our results indicate that in every case as $\lambda_{+}$ becomes large the counter-lasing phase must take hold at some point since the Raman transition described by $\lambda_{+}$ becomes much stronger than that of $\lambda_{-}$.

We remark that the ranges of $\lambda_{\pm}$ in Fig. \ref{fig:CouplingPlane} are large for illustrative purposes to highlight the structure of the bifurcation curves. In regions where the couplings are large, the behavior of the LMG model may depart somewhat from the Dicke model when the cavity field no longer adiabatically follows the atomic dynamics. In Appendix B, we compare the results of the semiclassical LMG model with the full semiclassical Dicke model to show that, in the parameter regime we are concerned with in the following discussion, the adiabatic elimination procedure is valid.

\begin{figure*}[t]
	\centering
	\includegraphics[width=17.9cm]{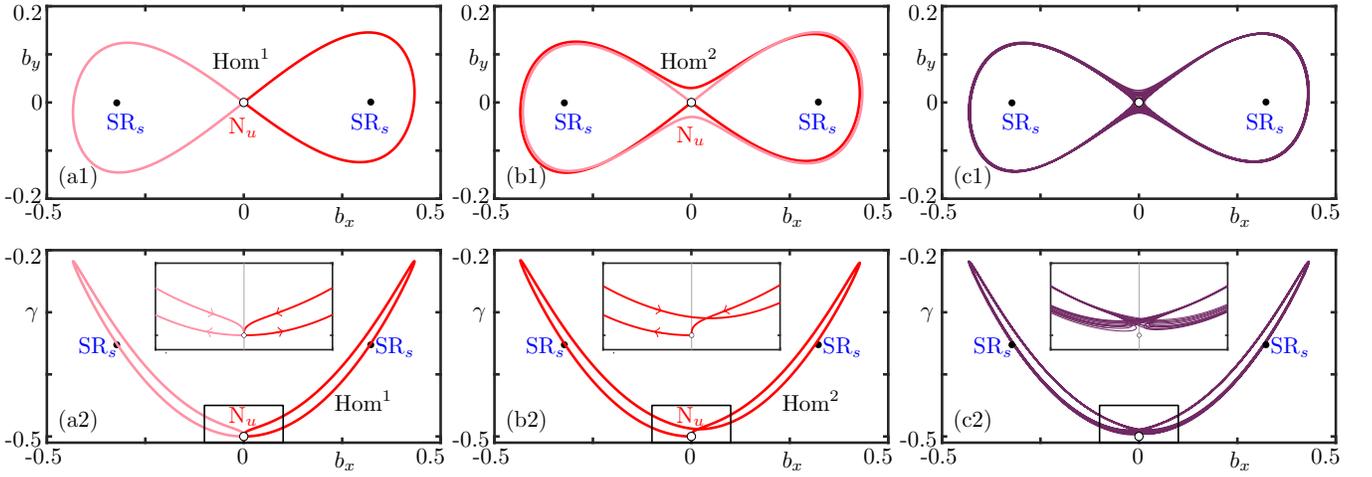}
	\caption{\label{fig:orbits} Phase space configuration of some relevant orbits at homoclinic bifurcations and inside the chaotic region. The pair of symmetry-related homoclinic orbits at the homoclinic bifurcation $\mathrm{Hom}^{1}$ is shown in panel (a1) in projection onto the $(b_{x},b_{y})$-plane and in panel (a2) in projection onto the $(b_{x},\gamma)$-plane. Panels (b1) and (b2) show the same projections of the pair of homoclinic orbits at $\mathrm{Hom}^{2}$, which feature an additional loop around the opposite superradiant equilibrium point; only one of the homoclinic orbits is shown in panel (b2). Panels (c1) and (c2) show in the same way the trajectory with initial condition $(b_{x},b_{y},\gamma)=(-0.035,-0.023,-0.495)$ which accumulates onto the chaotic attractor. In panels (a): $\lambda_{+}\approx 1.53208$, (b): $\lambda_{+}\approx 1.53286$, (c): $\lambda_{+}=1.53292$; other parameters are set as in Fig. \ref{fig:OneParBifDiag}.}
\end{figure*}

Once in the superradiant phase, upon increase of $\lambda_{+}$ the system enters the counter-lasing region \textbf{CL} through a thin transitional region where many bifurcations occur. This transition represents the mechanisms of how the superradiant pole-flip transition is perturbed with the addition of atomic dissipation, where the inverted normal phase is no longer viable. In contrast to the normal pole-flip transition, which goes from a transition involving infinitely many periodic orbits to a Hopf bifurcation when atomic dissipation is included, the superradiant pole-flip transition is made substantially more complex with the inclusion of atomic dissipation.

\begin{figure}[b]
	\centering
	\includegraphics[width=7cm]{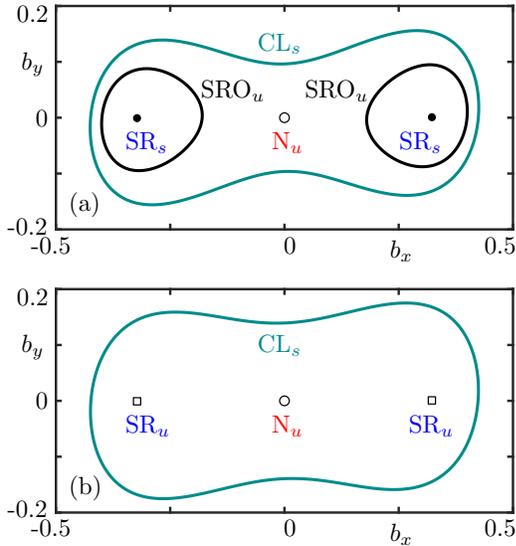}
	\caption{\label{fig:orbits2} Phase space configurations after the inverse period-doubling cascade (a) and after the subcritical Hopf bifurcation (b). For (a): $\lambda_{+} = 1.537$, for (b):$\lambda_{+}=1.54$. Other parameters are as in Fig. \ref{fig:OneParBifDiag}.}
\end{figure}

\subsubsection{One-Parameter Slice Through the Transitional Region}

To investigate the series of bifurcations that lead to the emergence of the counter-lasing phase inside the transitional region, we take a one-parameter slice through the phase diagram in Fig. \ref{fig:CouplingPlane} along the vertical line for $\lambda_{-} = 1.5$. The bifurcation diagram illustrating the transition as $\lambda_{+}$ is varied along this slice is shown in Fig.~\ref{fig:OneParBifDiag}(a); it has been obtained by numerical continuation of equilibria and periodic orbits with the program \textsc{auto-07p} \cite{AUTO1,AUTO2}. Due to the system's $\mathbb{Z}_{2}$ equivariance, this bifurcation diagram is symmetric about the horizontal axis, but for clarity we are only showing the upper half. The system starts for $\lambda_{+}$ small in the normal phase $\mathrm{N}_s$; here the subscript denotes stability: $s$ for stable, $u$ for unstable. As $\lambda_{+}$ is increased the system undergoes the supercritical pitchfork bifurcation $\mathrm{P}_\mathrm{spr}$ from which two stable superradiant equilibria $\mathrm{SR}_s$ emerge, and the normal phase becomes unstable, now labelled $\mathrm{N}_{u}$. As $\lambda_{+}$ is further increased, the equilibria $\mathrm{SR}_s$ undergo a pair of (subcritical) Hopf bifurcations at $\mathrm{H}_\mathrm{sub}$, where they bifurcate with a pair of unstable superradiant periodic orbits $\mathrm{SRO}_u$. Here, the superradiant equilibria become unstable ($\mathrm{SR}_u$); they disappear, as $\lambda_{+}$ is increased further, in a subcritical pitchfork bifurcation $\mathrm{P}_\mathrm{sub}$ where they meet the normal phase equilibrium $\mathrm{N}_u$.

Also shown in Fig. \ref{fig:OneParBifDiag}(a) is a branch of stable counter-lasing periodic orbits $\mathrm{CL}_{s}$ emerging from a small transitional region where they undergo interactions with the branch of superradiant periodic orbits $\mathrm{SRO}_{u}$. A magnification of these branches is shown in Fig. \ref{fig:OneParBifDiag}(b). Upon further enlargement in Fig. \ref{fig:OneParBifDiag}(c), we can glean some information on how this transitional region is organized. The beginning of the transition is marked by the creation of two branches of (saddle) periodic orbits, $\mathrm{SRO}_u$ and $\mathrm{CL}_u$, in a homoclinic bifurcation labelled $\mathrm{Hom}^1$, where there is a pair of homoclinic orbits to the normal phase equilibrium $\mathrm{N}_{u}$. As $\lambda_{+}$ is increased, the (unstable) counter-lasing branch of periodic orbits $\mathrm{CL}_u$ undergoes a pitchfork bifurcation of periodic orbits (PPO) upon collision with another branch of (stable, asymmetric) counter-lasing periodic orbits, $\mathrm{CL}_{s}^{\mathrm{asym}}$. This bifurcation renders the symmetric counter-lasing branch stable ($\mathrm{CL}_{s}$).

The pair of homoclinic orbits associated with the homoclinic bifurcation $\mathrm{Hom}^1$ responsible for creating the periodic orbits $\mathrm{SRO}_u$ and $\mathrm{CL}_u$ are shown projected onto the $(b_{x},b_{y})$-plane in Fig. \ref{fig:orbits}(a1) and the $(b_{x},\gamma)$-plane in Fig. \ref{fig:orbits}(a2). The two homoclinic orbits are symmetric counterparts under $\mathcal{T}$, and are homoclinic to the real saddle $\mathrm{N}_u$, which has one unstable and two stable directions. In the projection in Fig. \ref{fig:orbits}(a1), the pair of orbits forms a figure eight, while their configuration in three-dimensional space is that of an \emph{expanding butterfly} \cite{homburg_homoclinic_2010,shilnikov_bifurcations_2002}. Let $v_s$ and $v_u$ be the leading stable and unstable eigenvalues of the Jacobian matrix evaluated at $\mathrm{N}_u$, respectively, with corresponding eigenvectors $\vec{y}_s$ and $\vec{y}_u$. The butterfly configuration of homoclinic orbits occurs when both homoclinic orbits approach $\mathrm{N}_u$ from the same side along the direction given by $\vec{y}_s$ (the positive side of the $\gamma$-axis), but leave alongside opposite directions, $\vec{y}_u$ and $-\vec{y}_u$. This is illustrated in Fig. \ref{fig:orbits}(a2), where it can be seen that the pair of homoclinic orbits actually approach $\mathrm{N}_u$ from the same direction along the $\gamma$-axis. This configuration is referred to as expanding when $|v_{u}| > |v_{s}|$ \cite{shilnikov_bifurcations_2002}.

Figure~\ref{fig:OneParBifDiag}(d) shows an additional magnification of Fig.~\ref{fig:OneParBifDiag}(c) near the branch $\mathrm{CL}_{s}^{\mathrm{asym}}$. Here there are infinitely many more homoclinic bifurcations $\mathrm{Hom}^{2^k},\ k = 2,3,\cdots$, labelled by the number of loops $2^k$, which each create a pair of saddle periodic orbits. This leads to the formation of a chaotic attractor upon their accumulation, shown in Fig. \ref{fig:orbits}(c). As $\lambda_{+}$ is increased further, this chaotic attractor disappears in a period-doubling cascade (with period-doubling bifurcations $\mathrm{PD}_{k},\ k = 1,2,\cdots$). In the process the saddle periodic orbits coalesce into the branch of periodic orbits $\mathrm{CL}_{s}^{\mathrm{asym}}$, which then collides with the branch $\mathrm{CL}_u$ in the bifurcation PPO to create the stable counter-lasing branch $\mathrm{CL}_s$.

The homoclinic orbits at the second homoclinic bifurcation $\mathrm{Hom}^2$ are shown in Fig. \ref{fig:orbits}(b1) in the $(b_x,b_y)$-plane, and only one of them is shown in the $(b_x,\gamma)$-plane in Fig. \ref{fig:orbits}(b2). This bifurcation occurs along the branch $\mathrm{CL}^{\mathrm{asym}}$ in Fig. \ref{fig:OneParBifDiag}(d), which is formed at the pitchfork bifurcation of periodic orbits PPO and becomes unstable at the first period-doubling bifurcation $\mathrm{PD}_{1}$. Again, there are two coexistent homoclinic orbits in an expanding butterfly configuration, but now each features an additional loop in Fig. \ref{fig:orbits}(b1) around the symmetric counterpart of the superradiant equilibrium point $\mathrm{SR}_s$. In the same fashion, the homoclinic orbits at the subsequent homoclinic bifurcation $\mathrm{Hom}^{2^k},\ k = 2,3,\cdots$ feature $2^k-1$ additional loops around $\mathrm{SR}_s$. Figure \ref{fig:orbits}(c) shows the chaotic attractor that exists after the accumulation point of homoclinic bifurcations $\mathrm{Hom}^{2^k},\ k = 0,1,\cdots$, also in the same two projections. This attractor is localized in a small neighborhood of the union of the two initial homoclinic orbtis; compare with Fig. \ref{fig:orbits}(a). As such, it ``surrounds" the two (stable) superradiant equilibria $\mathrm{SR}_s$, and the chaotic trajectories switch irregularly between circling around (yet well away from) either of these stable equilibria. 

\begin{figure}[t]
	\centering
	\includegraphics[width=7.2cm]{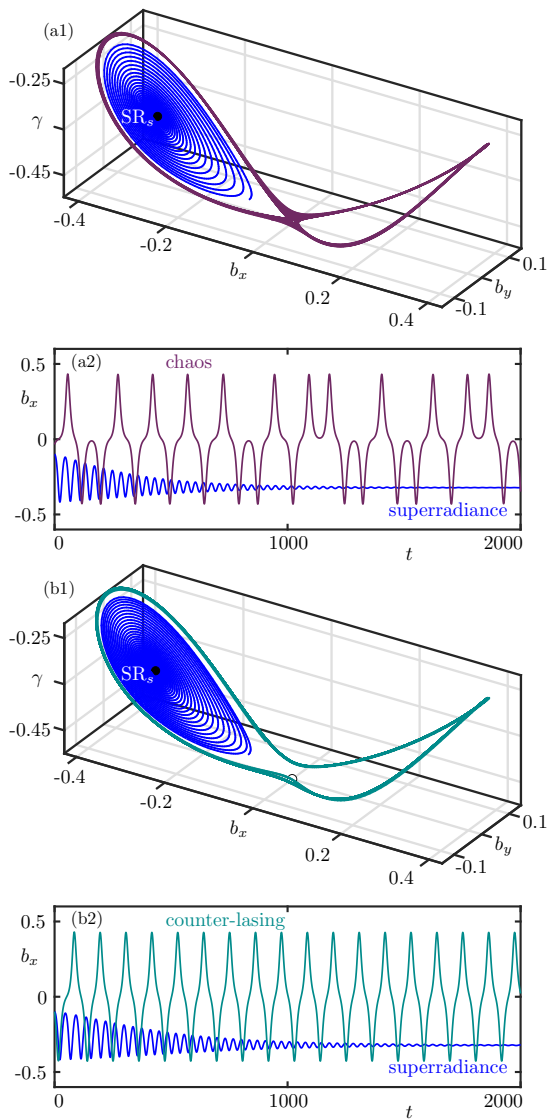}
	\caption{\label{fig:coexistence}Demonstration of phase coexistence in the semiclassical LMG model. Panel (a1) shows a trajectory with the same initial condition as Fig. \ref{fig:orbits}(d) which converges to the chaotic attractor (purple), while a trajectory (blue) with another initial condition $(b_{x},b_{y},\gamma) = (-0.1,0,-0.475)$ becomes superradiant. Panel (a2) shows the temporal trace of the two trajectories in $b_{x}$. Panel (b1) shows a trajectory with initial condition $(b_{x},b_{y},\gamma) = (-0.1,0.065,-0.475)$ which exhibits counter-lasing (green), and another trajectory which becomes superradiant (blue), with the same initial condition as in panel (a1). Panel (b2) also shows the temporal trace of these two trajectories in $b_{x}$. In panels (a1) and (a2), $\lambda_{+} = 1.53292$, and for panels (b1) and (b2) $\lambda_{+} = 1.534$. Other parameters are set to those in Fig. \ref{fig:OneParBifDiag}.}
\end{figure}

As $\lambda_{+}$ becomes larger, the chaotic attractor in Fig.~\ref{fig:orbits}(c) disappears in the (inverse) period-doubling cascade $\mathrm{PD}_{k},\ k = 1,2,\cdots$, shown in Fig. \ref{fig:OneParBifDiag}(d). The pair of periodic orbits remaining after this process are the (stable) asymmetric counter-lasing orbits $\mathrm{CL}_{s}^{\mathrm{asym}}$, which subsequently disappear in the pitchfork bifurcation of periodic orbits PPO; here they meet the unstable periodic orbit $\mathrm{CL}_u$, which then transitions to become stable ($\mathrm{CL}_s$). After this final transition in Fig. \ref{fig:OneParBifDiag}(d), the system now supports solutions that lase \emph{regularly}, as opposed to chaotically. Again, the superradiant periodic orbits $\mathrm{SRO}_u$ and $\mathrm{SR}_s$ remain unaffected. This configuration of equilibria and periodic orbits is shown in Fig. \ref{fig:orbits2}(a). Further along the bifurcaton diagram in Fig. \ref{fig:OneParBifDiag}(a), the periodic orbits $\mathrm{SRO}_u$ contract around the equilibria $\mathrm{SR}_s$ until the subcritical Hopf bifurcation $\mathrm{H}_{\mathrm{sub}}$ is reached, where the periodic orbits disappear and the equilibria become unstable ($\mathrm{SR}_{u}$). After this bifurcation, the counter-lasing periodic orbit $\mathrm{CL}_s$ is the only attractor, so the system is entirely in the counter-lasing phase. This situation is shown in Fig. \ref{fig:orbits2}(b).

\subsubsection{Coexistent Superradiance and Counter-Lasing}

The appearance of a chaotic attractor emerging from lasing states in the semiclassical model arising from the master equation (\ref{eqn:fullmastereqn}) was observed in Ref. \cite{kirton_superradiant_2018}. However, the situation presented here is slightly different, as the chaotic attractor here is associated with the appearance of \emph{counter-lasing} states; that is, the coherent driving terms occur with a greater strength than the incoherent pumping. As a result, in the chaotic regime, the system simultaneously supports chaotic counter-lasing and steady-state superradiance. The superradiant equilibria $\mathrm{SR}_s$ in Fig. \ref{fig:OneParBifDiag}(a) remain stable until the Hopf bifurcation $\mathrm{H}_{\mathrm{sub}}$, so there is a small region of overlap leading to phase coexistence of chaotic counter-lasing and superradiance. 

Figure \ref{fig:coexistence}(a1) illustrates the dynamics in this region in phase space. Here, we show two trajectories:
\begin{itemize}[itemsep=-5mm]
	\item one which accumulates onto the chaotic attractor, which lies beyond, yet surrounds the basins of attraction of the two superradiant equilibria $\mathrm{SR}_s$, \\
	\item and another, which approaches a superradiant state.
\end{itemize} 
The dramatic difference in the behavior generated by the coexistence of these different types of attractors is highlighted in Fig.~\ref{fig:coexistence}(a2), which shows the temporal trace of $b_{x}$ of the two trajectories shown in panel (a1). The trajectory on the chaotic attractor exhibits an irregular and nonrepeating dynamics with intermittent transitions between larger oscillations around either of the superradiant states, i.e., the trajectory chaotically switches between positive and negative $b_{x}$. This trajectory corresponds to the atomic ensemble lasing chaotically. Meanwhile, the trajectory in the basin of the superradiant equilibrium simply decays into the steady state. We have observed qualitatively that the basin of attraction of the chaotic attractor is quite small and includes the normal phase equilibrium $\mathrm{N}_u$. In particular, this means that trajectories with initial conditions sufficiently close to $\mathrm{N}_u$ accumulate onto the chaotic attractor and begin to lase chaotically. On the other hand, initial conditions with higher levels of excitation, that is, further away from $\mathrm{N}_u$, cause the atoms to begin to emit in a superradiant fashion, and eventually, regularly.
	
In the parameter regime after the (inverse) period-doubling cascade $\mathrm{PD}_{k},\ k = 1,2,\cdots$ and the pitchfork bifurcation of periodic orbits PPO, the chaotic attractor disappears and is replaced by a (symmetric under $\mathcal{T}$) stable counter-lasing periodic orbit $\mathrm{CL}_s$. As mentioned previously, the superradiant equilibria are unaffected by this transition. Thus, before the Hopf bifurcation $\mathrm{H}_{\mathrm{sub}}$, the system features phase coexistence of superradiance and regular counter-lasing. Figure \ref{fig:coexistence}(b1) shows two trajectories corresponding to the simultaneously existing superradiant and counter-lasing phases. Temporal traces of the trajectories' evolution are shown in Fig. \ref{fig:coexistence}(b2). Similarly to the case discussed above, initial conditions near $\mathrm{N}_u$ lead to periodic counter-lasing, while for most initial conditions atomic ensembles are emitting in a superradiant manner.

\subsection{Global Bifurcations via Kneading Sequences}

In order to investigate the dynamics leading to the establishment of chaos in more detail, we now turn our attention to characterizing relevant trajectories of Eqs.~(\ref{eqn:ODEs1})--(\ref{eqn:ODEs2}) with the use of parameter sweeping techniques based on the calculation of a \emph{kneading sequence}. Since their inception in Ref.~\cite{barrio_kneadings_2012}, these techniques have been used successfully to characterize the complicated arrangement of homoclinic (and other global) bifurcations in parameter space close to special organizing bifurcation points \cite{barrio_kneadings_2012,xing_symbolic_2014,giraldo_chaotic_2021}.

The kneading sequence encodes how a particular trajectory explores phase space. In particular, this facilitates an understanding of how complex switching patterns change as parameters are varied, thus, elucidating complex bifurcation orderings. To characterize kneading sequences, we will also make use of a mapping from the space of kneading sequences to the interval $[0,1]$, known as the \emph{kneading invariant} $K$. Thus, by sweeping $\lambda_{+}$ while holding other parameters constant, we compute (an approximation of) the kneading invariant $K$ for each $\lambda_{+}$ to show how the transitional region is organized. In doing so, we discover a complex recursive fractal structure, arranged around various global bifurcations, that gives insight into the emergence of complicated dynamics in this model.

To compute the kneading sequence, we consider the \emph{unstable manifold} $W^{u}(\mathrm{N}_{u})$ of the saddle equilibrium $\mathrm{N}_{u}$, which is the set of all points in phase space that converge towards $\mathrm{N}_{u}$ in backwards time, that is,
\begin{equation}
W^{u}(\mathrm{N}_{u}) = \bigg\{ x\in\mathbb{R}^{3} \bigg| \lim_{t\rightarrow-\infty}\varphi_{t}x = \mathrm{N}_{u} \bigg\},
\end{equation}
where $\varphi_{t}$ is the flow or time-evolution operator of Eqs.~(\ref{eqn:ODEs1})--(\ref{eqn:ODEs2}). In the parameter regime we are considering here, the unstable manifold $W^{u}(\mathrm{N}_{u})$ is one-dimensional, and changes in its kneading sequence allows one to study global bifurcations \cite{barrio_kneadings_2012} as $\lambda_{+}$ is varied. The manifold $W^{u}(\mathrm{N}_{u})$ has two branches, $W^{u}_{+}(\mathrm{N}_{u})$ and $W^{u}_{-}(\mathrm{N}_{u})$, which leave $\mathrm{N}_u$ tangent to the unstable eigenvector $\vec{y}_{u}$  in the positive and negative direction, respectively. As the normal phase equilibrium $\mathrm{N}_{u}$ lies within the parity symmetry subspace, the two branches of the unstable manifold are each other's symmetric counterparts; that is, $\mathcal{T}[W^{u}_{+}(\mathrm{N}_{u})] = W^{u}_{-}(\mathrm{N}_{u})$. As such, we only need to consider a single branch of the manifold to fully categorize the dynamics.

\begin{figure}[t]
	\centering
	\includegraphics[width=8.3cm]{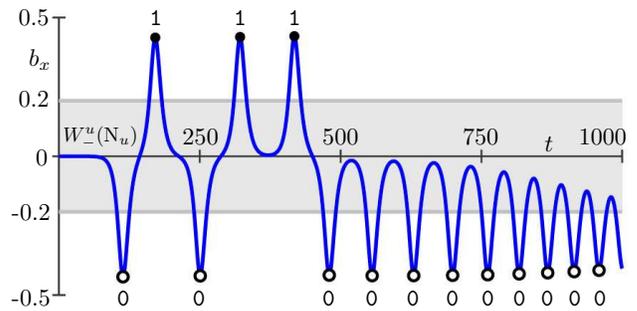}
	\caption{\label{fig:kneading_example} Kneading sequence $S = (\zero\one\zero\one\one\bar{\zero})$ generated by $W^{u}_{-}(\mathrm{N}_{u})$. Each negative minimum or positive maximum of $b_{x}$ outside of the exclusion zone $(-0.2,0.2)$ creates a symbol 0 or 1, respectively. Here, $\lambda_{+} = 1.532895$ and other parameters are set to those in Fig. \ref{fig:OneParBifDiag}.}
\end{figure}

We consider here the kneading sequence defined by $W^{u}_{-}(\mathrm{N}_{u})$. More specifically, we label its successive minima in $b_x$ with $b_x < 0$  with the symbols $\zero$, and maxima in $b_x$ with $b_x > 0$ with the symbol $\one$. To avoid misidentifying minima and maxima due to small oscillations near $\mathrm{N}_u$, we only consider those that fall outside of an \emph{exclusion zone} $-0.2 < b_{x} < 0.2$. This procedure defines the symbolic sequence of $\zero$s and $\one$s generated by $W^{u}_{-}(\mathrm{N}_{u})$, known as the \emph{kneading sequence}. It is unique for every point in parameter space and encodes the order in which $W^{u}_{-}(\mathrm{N}_{u})$ visits the two regions of phase space where $b_{x}$ is either negative or positive. Each kneading sequence is denoted $S = (a_{1}a_{2}a_{3}\cdots)$, where $a_{k}\in\{\zero,\one\}$ for $k\in\mathbb{N}$, and is a member of the set of all binary sequences, i.e., $S \in \{\zero,\one\}^{\mathbb{N}}$. Kneading sequences that end in a repeating infinite string of symbols will be denoted with an overbar, e.g., $\bar{\zero} \equiv \zero\zero\zero \cdots$ and $\overline{\zero\one} \equiv \zero\one\zero\one\zero\one\cdots$. Due to $\mathbb{Z}_{2}$ equivariance, the kneading sequences generated by either branch $W^{u}_{-}(\mathrm{N}_{u})$ or $W^{u}_{+}(\mathrm{N}_{u})$ are mapped to one another in the following way: for the kneading sequence $S = (a_1 a_2 a_3 \cdots)$ of $W^{u}_{-}(\mathrm{N}_u)$ the associated kneading sequence of $W^{u}_{+}(\mathrm{N}_u)$ takes the form 
\begin{equation}
\neg S = (\neg a_{1}\neg a_{2}\neg a_{3}\cdots),
\end{equation}
where $\neg$ is Boolean negation, that is,
\begin{equation}
\neg a_{k} = \begin{cases}
	\one, & \mathrm{if}\  a_{k} = \zero, \\
	\zero, & \mathrm{if}\  a_{k} = \one.
\end{cases}
\end{equation}
As an illustration, Fig. \ref{fig:kneading_example} shows the example where the unstable manifold generates the kneading sequence $S = (\zero\one\zero\one\one\bar{\zero})$, which we also write simply as $\zero\one\zero\one\one\bar{\zero}$ for notational convenience.

The key point here is that the kneading sequence of $W^{u}_{-}(\mathrm{N}_{u})$ is a topological invariant that allows one to identify global bifurcations \cite{barrio_kneadings_2012}. This means that the kneading sequence is constant in open regions of parameter space and, if it is different at two parameter values, then a homoclinic or heteroclinic bifurcation must have occurred. Thus, as $\lambda_{+}$ is varied, the kneading sequence of $W^{u}_{-}(\mathrm{N}_{u})$ can be used to identify the relevant bifurcations on the route to chaotic dynamics \cite{barrio_kneadings_2012,xing_symbolic_2014,giraldo_chaotic_2021}.

Typically (in open regions of parameter space), $W^{u}_{-}(\mathrm{N}_{u})$ has infinitely many extrema and so the kneading sequence indeed contains infinitely many symbols. However, homoclinic orbits, such as those shown in Fig.~\ref{fig:orbits}, make excursions exploring phase space only for some time before eventually closing back up at $\mathrm{N}_u$. We therefore describe a homoclinic orbit with a kneading sequence of only finitely many symbols, which are derived in the same way and encode how $W^{u}_{-}(\mathrm{N}_{u})$ visits the regions with negative and positive $b_x$, respectively, before returning to $\mathrm{N}_{u}$; in particular, the bit-length of such a finite kneading sequence is given by the number of loops of the homoclinic orbit. Hence, the finite kneading sequence of the homoclinic orbit formed by $W^{u}_{-}(\mathrm{N}_{u})$ in Fig.~\ref{fig:orbits}(a) is $\zero$ and that in Fig.~\ref{fig:orbits}(b) is $\zero\one$; their symmetric counterparts formed by $W^{u}_{+}(\mathrm{N}_{u})$ have kneading sequences $\one$ and $\one\zero$, respectively.

To facilitate an analysis of the kneading sequences, we consider the \emph{kneading invariant}
\begin{equation}
	K = \sum_{k=1}^{\infty} \frac{a_k}{2^k} \in [0,1],
\end{equation} 
which assigns a real number to each kneading sequence. In this way, performing a parameter sweep across $\lambda_{+}$ and computing the manifold $W_{-}^{u}(\mathrm{N}_{u})$, we can associate a kneading invariant $K$ for each value of $\lambda_{+}$ and identify global bifurcations for values where $K$ changes. Note that $K$ is not a one-to-one mapping, since the two kneading sequences
\begin{equation*}
	a_{1}a_{2}\cdots a_{k}\one\bar{\zero}\ \text{and}\ a_{1}a_{2}\cdots a_{k}\zero\bar{\one}
\end{equation*}
have the same kneading invariant $K$. Similarly, the finite kneading invariant of a homoclinic orbit is equal to the kneading invariant of the same sequence followed by an infinite string of $\zero$s.

In practice we can only consider kneading invariants arising from sequences of finitely many computed symbols, so we introduce the kneading invariant 
\begin{equation}
K_{n} = \sum_{k=1}^{n} \frac{a_k}{2^k}
\end{equation}
of sequences of bit-length $n$, which we use as an approximation of the full kneading invariant $K$. Note that $K_n$ divides the interval $[0,1]$ into $2^n$ subintervals, whose values of $K_n$ differ by at least $2^{-n}$; for our computations we use $n = 12$ throughout.


\begin{figure}[t!]
	\centering
	\includegraphics[width=8.6cm]{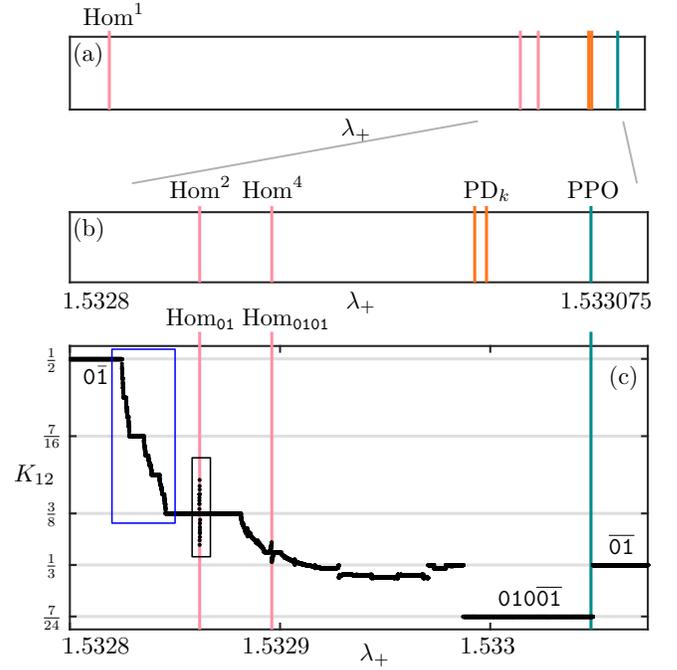}
	\caption{\label{fig:kneading1} Parameter sweep through $\lambda_{+}$ of the 12-symbol kneading invariant $K_{12}$. Panel (a) shows the parameter values for the bifurcations of Fig. \ref{fig:OneParBifDiag}(d), panel (b) is a magnification of the relevant $\lambda_{+}$-range, and  panel (c) shows the associated kneading invariant $K_{12}$, where the two frames indicate parts that are enlarged in Fig. \ref{fig:kneading2}. Other parameters are as in Fig.~\ref{fig:OneParBifDiag}.}
\end{figure}

\begin{figure}[t!]
	\centering
	\includegraphics[width=8.6cm]{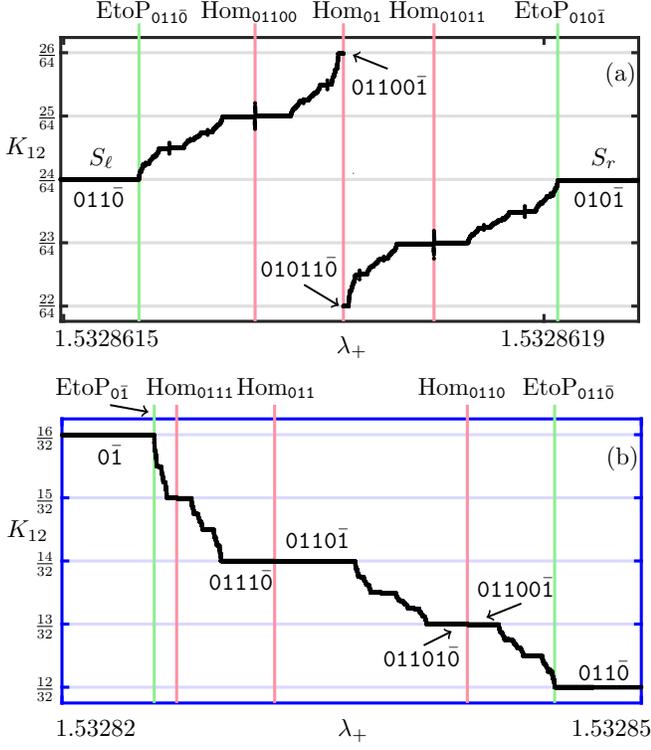}
	\caption{\label{fig:kneading2} Magnifications of the kneading invariant sweep in the two frames of Fig. \ref{fig:kneading1}(c). Panel (a) shows the structure near the homoclinic $\mathrm{Hom}_{\zero\one}$, and panel (b) shows the staircase formed between the EtoP connections $\mathrm{EtoP}_{\zero\bar{\one}}$ and $\mathrm{EtoP}_{\zero\one\one\bar{\zero}}$.}
\end{figure}

To see how the transitional region in Fig. \ref{fig:OneParBifDiag} is organized, we now execute a parameter sweep that determines the kneading invariant $K_{12}$ as a function of $\lambda_{+}$. To indicate the $\lambda_{+}$-interval of the sweep, we plot in Fig.~\ref{fig:kneading1}(a) the values of $\lambda_{+}$ for which there are homoclinic, period-doubling, and pitchfork bifurcations of periodic orbits already found by numerical continuation; see Fig.~\ref{fig:OneParBifDiag}. An enlargement of the relevant range between the bifurcations $\mathrm{Hom}^{2}$ and $\mathrm{PPO}$ is shown in Fig.~\ref{fig:kneading1}(b). Over a fine grid of this $\lambda_{+}$-interval, we compute, for each value of $\lambda_{+}$ in the sweep, the one-dimensional unstable manifold $W^{u}_{-}(\mathrm{N}_{u})$ accurately up to a length that allows us to determine the associated 12-symbol initial part of the kneading sequence; this is achieved with the software package \textsc{tides} \cite{abad_algorithm_2012}, which uses a Taylor series method for high-precision numerical integration of ordinary differential equations.

Figure \ref{fig:kneading1}(c) shows the resulting plot of the kneading invariant $K_{12}$, and we first discuss its overall features. At the start of the sweep there is an interval where the kneading sequence is $\zero\bar{\one}$ with a kneading invariant $\frac{1}{2}$, and at the end of the sweep there is an interval where it is $\overline{\zero\one}$ with a kneading invariant $\frac{1}{3}$. In between, the kneading invariant is constant on smaller and smaller intervals and undergoes a remarkably complex series of changes that correspond to the sequence of global bifurcations that give rise to homoclinic chaos \cite{barrio_kneadings_2012}. On the left of Fig. \ref{fig:kneading1}(c) (in the larger frame) there is a descending ``staircase" structure; it will be discussed later in the section. Following this, as $\lambda_{+}$ increases further, we find a ``plateau'' where the kneading invariant is constant over an interval, followed by a sharp, spiked increase and decrease of $K_{12}$ (in the smaller frame). It is centered around a homoclinic bifurcation $\mathrm{Hom}^{2}$, which we now refer to by its finite kneading sequence as $\mathrm{Hom}_{\zero\one}$; similarly, we refer to any homoclinic bifurcation as $\mathrm{Hom}_{S}$ from now on, where $S$ is its finite kneading sequence. Notice that there is another spiked increase and decrease of $K_{12}$ near $\mathrm{Hom}^{4} = \mathrm{Hom}_{\zero\one\zero\one}$. The nature of this large local change of $K_{12}$ will be discussed below. As $\lambda_{+}$ is increased further, near the central range in Fig. \ref{fig:kneading1}(c) where we expect a chaotic attractor to appear, $K_{12}$ flattens out; this is due to changes in the kneading sequences occurring only at later and later symbols, with correspondingly smaller changes to the kneading invariant. As $\lambda_{+}$ is further increased, the kneading invariant makes two large jumps. First down to $\frac{7}{24}$  and an interval where the kneading sequence is $\zero\one\zero\overline{\zero\one}$, after which there is a second jump at the pitchfork bifurcation of periodic orbits PPO, up to $\frac{1}{3}$ and the final interval with kneading sequence $\overline{\zero\one}$, which corresponds to the stabilization of the symmetric counter-lasing phase.

Figure~\ref{fig:kneading2} shows magnifications of the kneading invariant plot of Fig.~\ref{fig:kneading1}(c) in the two highlighted frames. We first turn our attention to the plateau and the spike near the homoclinic bifurcation $\mathrm{Hom}_{\zero\one}$, which is shown over an enlarged $\lambda_{+}$-range in Fig.~\ref{fig:kneading2}(a). This closer inspection reveals that the sharp spike in the kneading invariant is a smaller structure with additional plateaux with spikes at further homoclinic bifurcations, thus, revealing an interesting type of self-similarity. Notice that there are two kneading sequences associated with each plateau, one to the left of the homoclinic bifurcation, and one to the right. Specifically, the start and end of the main plateau in Fig.~\ref{fig:kneading2}(a) is generated by the kneading sequences $S_\ell = (\zero\one\one\bar{\zero})$ and $S_r = (\zero\one\zero\bar{\one})$, respectively, which both have the same kneading invariant $K = \frac{3}{8}$ (but differ by $2^{-12}$ in $K_{12}$). The `spike' is identified here as consisting of two additional staircase structures, similar to the one seen in Fig. \ref{fig:kneading1}(c) --- an increasing one to the left of $\mathrm{Hom}_{\zero\one}$ and a decreasing one to the right of $\mathrm{Hom}_{\zero\one}$, which meet at the central organizing homoclinic bifurcation $\mathrm{Hom}_{\zero\one}$ with a discontinuity in the kneading invariant $K$. Each staircase structure in Fig.~\ref{fig:kneading2}(a) has its own central, largest plateau with a spike at the homoclinic bifurcations $\mathrm{Hom}_{\zero\one\one\zero\zero}$ and $\mathrm{Hom}_{\zero\one\zero\one\one}$, respectively, and within each of the plateau of the staircase we find further staircases with infinitely more plateaux and staircases, and so on. 

The staircase structure we find highly resembles the Cantor function \cite{dovgoshey_cantor_2006} --- also known as the Cantor staircase or ``Devil's staircase" ---  which is a continuous and monotone (increasing or decreasing) function that is locally constant. However, the graph of $K$ we find here has the additional spikes in the middle of each plateau, so is neither monotone nor continuous and features additional self-similarity since each spike consists of two such staircase structures. We therefore refer to the graph of $K$ shown in Fig.~\ref{fig:kneading2}(a) as a \emph{spiked (Cantor) staircase}, and we now proceed to consider in more detail how it is organized recursively by sequences of homoclinic orbits. 

Firstly, all kneading sequences of the spike in Fig.~\ref{fig:kneading2}(a) start with the common first two symbols $\zero\one$ of $S_\ell$ and $S_r$, the kneading sequences at the plateau we start with. Moreover, for the entire left spiked staircase for $\lambda_{+} < \lambda_{+}^{\mathrm{Hom_{\zero\one}}}$ this is followed by the symbol $\one$, which means that the kneading invariant $K$ is larger than the value of the plateau itself. In contrast, for the right spiked staircase with $\lambda_{+} > \lambda_{+}^{\mathrm{Hom}_{\zero\one}}$ any kneading sequence always starts with $\zero\one\zero$, meaning the kneading invariant is always less than the value of the plateau. This combinatorial structure is repeated at each sub-plateau, as we checked for the respective largest one, where we find spikes (that is, additionally left and right spiked staircases) near the homoclinic orbits with the finite kneading sequences $\zero\one\one\zero\zero$ and $\zero\one\zero\one\one$. We remark that we have identified these and other additional homoclinic orbits by computing them with a Lin's method approach \cite{lin_using_1990}; see Appendix~\ref{app:lin} for details. 

As a general observation, we find the following for each homoclinic bifurcation $\mathrm{Hom}_{a_1 a_2 \cdots a_k}$. For each plateau of its left spiked staircase with kneading sequence 
\begin{equation*}
S_\ell = (a_1 a_2 \cdots a_k \, \one \, a_{k+2} \cdots)
\end{equation*}
there is a corresponding plateau on its right spiked staircase with the kneading sequence
\begin{equation*}
S_r = (a_1 a_2 \cdots a_k \ \neg [\one \, a_{k+2}\cdots]).
\end{equation*}
This relationship of negating/inverting the symbols after the common prefix determines a map from the left to the right spiked staircase near each homoclinic bifurcation and, thus, explains the observed symmetry in Fig.~\ref{fig:kneading2}(a). As a concrete example, note that 
the kneading sequence of the main plateau to the left of the homoclinic bifurcation $\mathrm{Hom}_{\zero\one}$ is $S_\ell = (\zero\one\one\bar{\zero})$, which hence maps to its right counterpart with kneading sequance $S_r = (\zero\one\neg[\one\bar{\zero}])= (\zero\one\zero\bar{\one})$. Note further, that this observation also explains the self-similar structure of the graph of $K$: apart from the common prefix and the fact that $a_{k+1} = \one$ for any plateau of the left spiked staircase, there is no restriction on the symbols $a_{i} \in \{\zero,\one\}$ for $i \geq k+2$. Hence, at every level, there is a full binary tree of further spiked plateaux of $K$, each with its own central homoclinic bifurcation.

Figure \ref{fig:kneading2}(b) is an enlargement of the first part of the staircase structure in Fig. \ref{fig:kneading1}(c) between the initial plateau with kneading sequence $\zero\bar{\one}$ with $K = \frac{1}{2}$ and the next largest plateau with kneading sequence $\zero\one\one\bar{\zero}$ with $K = \frac{3}{2}$. At first glance, it looks like a regular descending Devil's staircase, but it does also feature spikes at every sub-plateau, the most prominent of which is the spike in Fig.~\ref{fig:kneading2}(a) we just discussed. Hence, it is also a spiked staircase, and we find that it features plateaux and associated homoclinic bifurcations with kneading sequences composed of all sequences arising from the binary tree starting with the symbols $\zero\one\one$. Specifically, the largest plateau in panel (b) is centered around $\mathrm{Hom}_{\zero\one\one}$, with the next two largest inheriting the prefix and centered around $\mathrm{Hom}_{\zero\one\one\zero}$ and $\mathrm{Hom}_{\zero\one\one\one}$, which are homoclinic orbits that we also identified numerically (see again Appendix~\ref{app:lin}). While the spikes in panel (b) near these homoclinic bifurcations are smaller (extend less in $K$ from the value of the respective plateau) compared to panel (a), each indeed consists of a left and a right spiked sub-staircase. 

It is clear from the above discussion that the number of homoclinic bifurcations $\mathrm{Hom}_{S}$, as given by their finite symbol sequences $S$, is countably infinite. In particular, this means that there must be accumulation points of homoclinic bifurcations in the finite $\lambda_+$-range where they occur. As we have checked and show now, the two boundaries of each plateau, that is, the left and right end points of a spike, are limits of specific sequences of homoclinic bifurcations, namely those with more and more symbols $\zero$ and $\one$ after the common prefix, respectively. For example, in Fig. \ref{fig:kneading2}(a) the sequence  $\mathrm{Hom}_{\zero\one\one}$, $\mathrm{Hom}_{\zero\one\one\zero}$, $\mathrm{Hom}_{\zero\one\one\zero\zero}$, $\cdots$ converges to the end point of the part of the plateau with kneading sequence $S_\ell = (\zero\one\one\bar{\zero})$, and the sequence $\mathrm{Hom}_{\zero\one\zero}$, $\mathrm{Hom}_{\zero\one\zero\one}$, $\mathrm{Hom}_{\zero\one\zero\one\one}$, $\cdots$ converges to the end point of the plateau with kneading sequence $S_r = (
\zero\one\zero\bar{\one})$. Similarly, in Fig. \ref{fig:kneading2}(b) the sequence  $\mathrm{Hom}_{\zero\one\one}$, $\mathrm{Hom}_{\zero\one\one\one}$, $\mathrm{Hom}_{\zero\one\one\one\one}$, $\cdots$ limits on the plateau with kneading sequence $\zero\bar{\one}$ at $K = \frac{1}{2}$, and the sequence $\mathrm{Hom}_{\zero\one\one}$, $\mathrm{Hom}_{\zero\one\one\zero}$, $\mathrm{Hom}_{\zero\one\one\zero\zero}$, $\cdots$ on that with kneading sequence $\zero\one\one\bar{\zero}$ at $K = \frac{3}{8}$.

It turns out that the limits of the respective sequences of homoclinic orbits exist as objects in phase space: they are heteroclinic connections from the saddle equilibrium $\mathrm{N}_u$ to either of the symmetric pair of saddle periodic orbits labelled $\mathrm{SRO}_u$ in Fig. \ref{fig:orbits2}(a). We refer to such connecting orbits as \emph{EtoP connections}, and they occur at isolated $\lambda_+$-values where the one-dimensional unstable manifold $W^{u}(\mathrm{N}_{u})$ lies in the two-dimensional stable manifold $W^{s}(\mathrm{SRO}_{u})$. We distinguish each limiting EtoP connection $\mathrm{EtoP}_S$ by its associated kneading sequence $S$, which either ends in $\bar{\zero}$ or $\bar{\one}$ when $W^{u}_-(\mathrm{N}_{u})$ accumulates on the periodic orbit with $b_x < 0$ or with $b_x > 0$, respectively. We remark that EtoP connections can also be identified reliably with a Lin's method approach \cite{lin_using_1990}; see Appendix~\ref{app:lin}. In this way, we found the limiting EtoP connections bounding the plateau with kneading sequences $S_\ell = (\zero\one\one\bar{\zero})$ and $S_r = (\zero\one\zero\bar{\one})$ and with kneading sequences $\zero\bar{\one}$ and $\zero\one\one\bar{\zero}$, the $\lambda_+$-values of which are indicated in panels (a) and (b) of Fig. \ref{fig:kneading2}.

\begin{figure}[h]
	\centering
	\includegraphics[width=8.6cm]{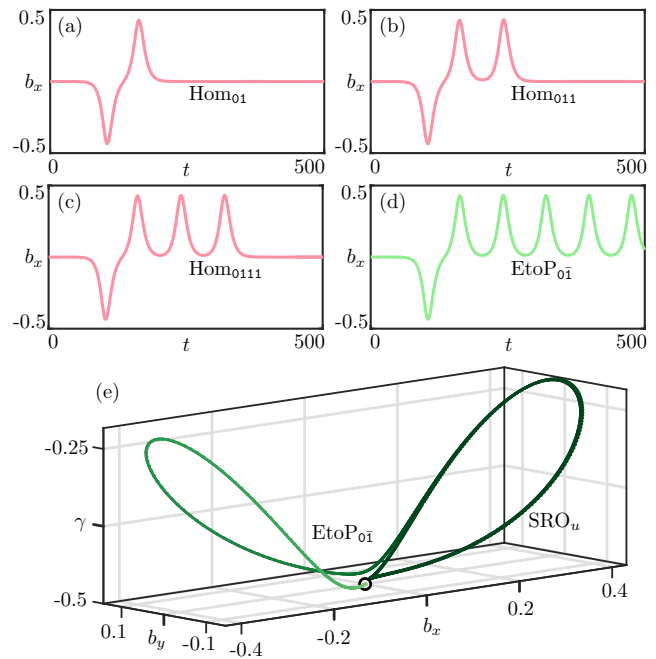}
	\caption{\label{fig:etop} Convergence of homoclinic orbits to an EtoP connection. Panels (a)--(c) show the temporal traces of the homoclinic orbits $\mathrm{Hom}_{01}$, $\mathrm{Hom}_{011}$ and $\mathrm{Hom}_{0111}$ as they accumulate on $\mathrm{EtoP}_{0\bar{1}}$, shown as a temporal trace in panel (d) and in $(b_x,b_y,\gamma)$-space  in panel (e).}
\end{figure}

The convergence of homoclinic orbits is illustrated in Fig. \ref{fig:etop} for the example of $\mathrm{EtoP}_{\zero\bar{\one}}$. As the temporal traces of $W^{u}_-(\mathrm{N}_{u})$ in panels (a)--(c) show, the homoclinic orbits $\mathrm{Hom}_{\zero\one}$, $\mathrm{Hom}_{\zero\one\one}$, and $\mathrm{Hom}_{\zero\one\one\one}$ each have one further excursion around the periodic orbit $\mathrm{SRO}_u$ with $b_x > 0$ before $W^{u}_-(\mathrm{N}_{u})$ returns to the saddle equilibrium $\mathrm{N}_u$. In the limit, $W^{u}_-(\mathrm{N}_{u})$ accumulates on $\mathrm{SRO}_u$, generating the EtoP connection $\mathrm{EtoP}_{\zero\bar{\one}}$, which is shown as a temporal profile in Fig. \ref{fig:etop}(d) and as a connecting orbit in $(b_x,b_y,\gamma)$-space in panel (e).

The limiting EtoP connections exist as part of the overall self-similar structure of the spiked staircase, where each plateau is bounded by the respective two EtoP connections. The tree of kneading sequences of a plateau and its spike is determined by its central homoclinic orbit $\mathrm{Hom}_{S}$ with given finite sequence $S$, which provides the prefix for all further homoclinic bifurcations and EtoP connections, as well as the associated plateaux. Note that plateaux with prefixes of large bit-lengths (before periodic repetition) correspond to trajectories in phase space that feature long chaotic transients before they settle down to either of the pair of superradiant equilibria $\mathrm{SR}_s$; see Fig.~\ref{fig:orbits2}(a). These chaotic transients may become arbitrarily long as part of the self-similar structure of the spiked Cantor staircase in Figs.~\ref{fig:kneading1} and~\ref{fig:kneading2}. We therefore suggest that in this range of $\lambda_+$-values prior to when the chaotic attractor emerges, Eqs.~(\ref{eqn:ODEs1})--(\ref{eqn:ODEs2}) displays \emph{preturbulence} --- in recognition of the analogy with the preturbulent regime of the Lorenz system prior to its transition to a chaotic attractor due to a single EtoP connection \cite{doedel_global_2015}.

%
%

\section{Conclusions}

We analyzed the nonlinear semiclassical physics arising from phase transitions leading to lasing and counter-lasing phases in a generalized Lipkin-Meskov-Glick model derived from a generalized Dicke model after adiabatic elimination in the dissipative limit. We first performed a study of the model in what we term the strong dissipative limit, where the dissipative terms of the quantum master equation greatly outweigh the unitary dynamics. In doing so, we have described the effect of adding spontaneous emission to the dynamics of a collective spin-flip transition. We call this the \emph{normal pole-flip transition}, and it manifests itself as a Hopf bifurcation to a counter-lasing periodic orbit. This can lead to resurgent superradiant pulses driven by spontaneous emission. We then presented a detailed analysis of the dynamics associated with the appearance of these counter-lasing periodic orbits, as well as how they can be reconciled with the inverted (spin-up) normal phase that appears in the limit of disappearing spontaneous emission. 

We then focussed on the transition to the counter-lasing phase once the system's parity symmetry is broken in a superradiant phase transition, as well as its relation to another collective spin-flip transition involving superradiant states, which we call the \emph{superradiant pole-flip transition}. We discovered a surprisingly complex sequence of bifurcations that organize this transition, which involves the establishment of a chaotic attractor. Specifically, we identified sequences of homoclinic bifurcations to the now unstable symmetric equilibrium, as well as sequences of connecting EtoP orbits from it to saddle periodic orbits, and showed that these are organized in an intricate structure we call a spiked Cantor staircase when one considers their associated topological quantity known as the kneading invariant. As we showed, this phenomenon entails long chaotic transients, which we characterized as a preturbulent regime that exists before the chaotic attractor emerges. 

Overall, this work demonstrates the appearance of highly complicated nonlinear dynamics from a dissipative many-body quantum optical model in the semiclassical limit. In this description, the dissipative phase transitions were shown to result from the interplay between the coherent processes that underpin superradiance and the incoherent processes that seed counter-lasing, and can be very complex in the transitional region where the two are in competition. 

A possible future direction of research that this work suggests is an analysis of the semiclassical dynamics of the full unbalanced Dicke model with atomic dissipation, outside of the dissipative limit where the use of the LMG model description is justified. Here, rather than three-dimensional, the semiclassical dynamics are actully five-dimensional by virtue of the dynamics of the light field no longer adiabatically following the atomic states. It would be interesting to investigate how, as the validity of the adiabatic elmination procedure is departed, the dynamics described in this paper are modified by the states of the light field lagging further and further behind the atomic states. Moreover, once entirely outside of the dissipative regime, there is the possibility of the appearance of additional complex dynamics and phase transition mechanisms between superradiance and counter-lasing in addition to those analyzed here. 

%
%

\appendix
\begin{widetext}
\section{Calculation of Local Bifurcation Conditions in the LMG Model}
\label{app:locbif}

The Jacobian matrix corresponding to Eqs. (\ref{eqn:ODEs1})--(\ref{eqn:ODEs2}) evaluated at the normal phase equilibium $b_{x} = b_{y} = 0$, $\gamma = \gamma_{\mathrm{eq}}$ is
\begin{equation}\label{eqn:JacobianLMG}
J = \begin{bmatrix}
2\eta\gamma_{\mathrm{eq}}(\lambda_{-}^2 - \lambda_{+}^2) - (\Gamma_\downarrow + \Gamma_\uparrow) & \omega_0 + 2\xi\gamma_{\mathrm{eq}}(\lambda_{-} - \lambda_{+})^2 & 0 \\
-\omega_0 - 2\xi\gamma_{\mathrm{eq}}(\lambda_{-}+\lambda_{+})^2 & 2\eta\gamma_{\mathrm{eq}}(\lambda_{-}^2 - \lambda_{+}^2) - (\Gamma_\downarrow + \Gamma_\uparrow) & 0 \\
0 & 0 & -2(\Gamma_\downarrow + \Gamma_\uparrow)
\end{bmatrix}.
\end{equation}

\subsection{Pitchfork Bifurcation Condition}

With the nonzero $\xi$-terms included, the eigenvalues of this matrix are complicated and are not of much help in determining the locations of the pitchfork or Hopf bifurcations in the $(\lambda_{-},\lambda_{+})$-plane. Instead, a necessary condition for a pitchfork bifurcation curve describing the breaking of the LMG model's $\mathbb{Z}_2$ symmetry is that
\begin{equation}\label{eqn:detJacobianLMG}
\det(J) = -2(\Gamma_\uparrow + \Gamma_\downarrow)\bigg[ (\Gamma_\downarrow + \Gamma_\uparrow)^2 + 4\gamma_\mathrm{eq}\eta(\Gamma_\downarrow+\Gamma_\uparrow)(\lambda_{+}^2-\lambda_{-}^2) + 4\gamma_\mathrm{eq}^2 (\eta^2 + \xi^2)(\lambda_{+}^2 - \lambda_{-}^2)^2 + 4\gamma_\mathrm{eq}\xi\omega_0(\lambda_{+}^2+\lambda_{-}^2)+\omega_0^2 \bigg] = 0.
\end{equation}
This equation is quadratic in $\lambda_{+}^2$, so an analytic solution is readily found and subsequently confirmed to be a pitchfork bifurcation. Asserting $\Gamma_\downarrow + \Gamma_\uparrow > 0$, this gives the locations of pitchfork bifurcation shown in Fig. \ref{fig:CouplingPlane}. 

\subsection{Saddle-node Bifurcation Condition}

The Jacobian matrix $J$ also yields a necessary condition for saddle-node bifurcations when it is evaluated at the superradiant equilibria (obtained with a computer algebra package), giving from $\det(J) = 0$ the equation
\begin{equation}
	2\eta\xi\omega_0(\Gamma_\downarrow + \Gamma_\uparrow)(\lambda_{-}^4 - \lambda_{+}^4) + \bigg[ \xi^2(\Gamma_\downarrow + \Gamma_\uparrow)^2 + \eta^2\omega_0^2 \bigg](\lambda_{-}^2 - \lambda_{+}^2)^2 - 4\xi^2\omega_0^2\lambda_{-}^2\lambda_{+}^2 = 0,
\end{equation}
which turns out to be linear in $\lambda_{-}$, yielding
\begin{equation} \label{eqn:SNlinear}
	\lambda_{+} = \mp\left\{ \frac{\sqrt{ \xi^2(\Gamma_\uparrow + \Gamma_\downarrow)^2 + \omega_0^2(\eta^2 + 2\xi^2) \pm  2\xi\omega_0\sqrt{ (\eta^2 + \xi^2)\big[ (\Gamma_\uparrow + \Gamma_\downarrow)^2 + \omega_0^2 \big] } }}{\xi(\Gamma_\uparrow + \Gamma_\downarrow) - \eta\omega_0}\right\}\ \lambda_{-}.
\end{equation}
We validated with numerical continuation that Eq. (\ref{eqn:SNlinear}) indeed gives the curves of saddle-node bifurcations of the superradiant equilibria.

\subsection{Hopf Bifurcation Condition}

A necessary condition for the Hopf bifurcation is formulated in terms of the \emph{bialternate product} $\odot$  of the Jacobian matrix $J$ with twice the identity $I$ \cite{kuznetsov_elements_2004} (evaluated at the normal phase equilibrium), which is
\begin{equation}
J\odot 2I = \begin{bmatrix}
4\eta\gamma_\mathrm{eq}(\lambda_{-}^2-\lambda_{+}^2) - 2(\Gamma_\uparrow+\Gamma_\downarrow) & 0 & 0 \\
0 & 2\eta\gamma_\mathrm{eq}(\lambda_{-}^2 - \lambda_{+}^2) - 3(\Gamma_\uparrow+\Gamma_\downarrow) & \omega_0 + 2\xi\gamma_\mathrm{eq}(\lambda_{-}-\lambda_{+})^2 \\
0 & -\omega_0 - 2\xi\gamma_\mathrm{eq}(\lambda_{-}+\lambda_{+})^2 & 2\eta\gamma_\mathrm{eq}(\lambda_{-}^2-\lambda_{+}^2)-3(\Gamma_\downarrow+\Gamma_\uparrow)
\end{bmatrix}.
\end{equation}
Its determinant is
\begin{equation}\label{eqn:bialtDet}
\det(J\odot 2I) = -2\bigg[ 2\eta\gamma_{\mathrm{eq}}(\lambda_{+}^2 - \lambda_{-}^2) + (\Gamma_\uparrow + \Gamma_\downarrow) \bigg]F(\lambda_{-},\lambda_{+}),
\end{equation}
where $F(\lambda_{-},\lambda_{+})$ is a quartic in $\lambda_{-}$ and $\lambda_{+}$. The eigenvalues of the matrix $J\odot 2I$ are the pairwise sums of the eigenvalues of $J$ \cite{kuznetsov_elements_2004}. Thus, there are two ways that $\det(J\odot 2I)=0$, which is the condition of interest: the matrix $J$ either has a complex conjugate pair of eigenvalues with zero real part (that is, a Hopf bifurcation), or it has a pair of real eigenvalues of $J$ that sum to zero. We identify the latter condition to correspond to $F(\lambda_{-},\lambda_{+}) = 0$, so the defining condition for the Hopf bifurcation is 
\begin{equation}\label{eqn:defcondHopf}
	2\eta\gamma_\mathrm{eq}(\lambda_{+}^2 - \lambda_{-}^2) + (\Gamma_\uparrow + \Gamma_\downarrow) = 0.
\end{equation}
This is exactly the condition (\ref{eqn:HopfCurves}) in the strong dissipative limit $\xi = 0$, which was again confirmed by numerical continuation. We remark that Eq.~(\ref{eqn:defcondHopf}) is a necessary condition, as the zeros of Eq.~(\ref{eqn:bialtDet}) can also be used to compute Hopf-degeneracies such as the pole-flip transition \cite{stitely_nonlinear_2020}.
\end{widetext}

\section{Validity of Adiabatic Elimination}
\label{app:adiabatic}

Here we will briefly discuss the validity of the adiabatic elimination process that leads to the semiclassical Lipkin-Meshkov-Glick (LMG) model [Eqs. (\ref{eqn:ODEs1})--(\ref{eqn:ODEs2})] in the parameter regime of Fig. \ref{fig:CouplingPlane}. For comparison, the semiclassical equations of motion arising from the Dicke model master equation (\ref{eqn:fullmastereqn}) are
\begin{align}
	\frac{d\alpha}{dt} ={}& -\kappa\alpha - i\omega\alpha - i\lambda_{-}\beta - i\lambda_{+}\beta^*, \label{eqn:dickeODEs1} \\
	\frac{d\beta}{dt} ={}& -i\omega_0 \beta + 2i\lambda_{-}\alpha\gamma + 2i\lambda_{+}\alpha^* \gamma - (\Gamma_\uparrow + \Gamma_\downarrow)\beta, \label{eqn:dickeODEs2} \\
	\frac{d\gamma}{dt} ={}& i\lambda_- (\alpha^* \beta - \alpha \beta^*) + i\lambda_{+}(\alpha\beta - \alpha^*\beta^*) \nonumber \\
	&- 2\Gamma_\downarrow \left( \frac{1}{2} + \gamma \right) + 2\Gamma_\uparrow\left( \frac{1}{2} - \gamma \right) \label{eqn:dickeODEs3},
\end{align}
where $\alpha = \braket{\hat{a}}/\sqrt{N}$. The adiabatic elimination procedure asserts that the cavity field amplitude $\alpha$ is effectively in the steady state ($d\alpha/dt = 0$) when the system is dissipative $\kappa \gg \omega, \omega_0,\lambda_-, \lambda_+$. The cavity field is then determined solely by the spin dynamics. 

\begin{figure}[t!]
	\centering
	\includegraphics[width=8.6cm]{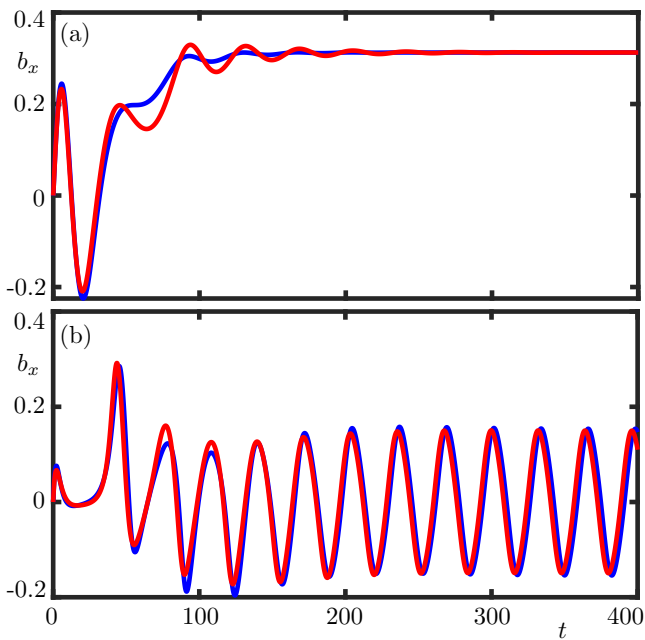}
	\caption{\label{fig:dicke_compare} Comparison of semiclassical dynamics of the Dicke model (blue) and the LMG model (red) in the superradiant phase (a) and the counter-lasing phase (b). Parameters are as in Fig. \ref{fig:OneParBifDiag}, with $\lambda_{+} = 1.5$ for (a) and $\lambda_{+} = 2$ for (b).}
\end{figure}

In Fig. \ref{fig:dicke_compare} we show the results of numerical integration of Eqs. (\ref{eqn:dickeODEs1})--(\ref{eqn:dickeODEs3}) and Eqs. (\ref{eqn:ODEs1})--(\ref{eqn:ODEs2}) for the same parameter values. Figure \ref{fig:dicke_compare}(a) shows the semiclassical Dicke and LMG model trajectories as they approach a superradiant equilibium point. The LMG model trajectory slightly underestimates the effective decay rate into the superradiant state and, as a result, converges to the state more slowly. However and crucially, the two trajectories reach the same steady-state. Figure \ref{fig:dicke_compare}(b) shows the trajectories converging to the counter-lasing phase. Again, the initial transients of the two trajectories differ slightly: the Dicke model trajectory accumulates a larger phase which causes the LMG model trajectory to slightly lag behind the Dicke model trajectory once the counter-lasing periodic orbit is reached. However, the periodic orbits of the two are essentially the same. Overall, there is qualitative agreement between the two models in this parameter regime, and good quantitative agreement of the long-term dynamics.

\section{Lin's Method}
\label{app:lin}

We now briefly outline the numerical implementation of \emph{Lin's method} \cite{lin_using_1990,krauskopf_lin_2008} we used to find homoclinic orbits and EtoP connections  at specific values of $\lambda_{+}$. It is based on the continuation of a suitable two-point boundary value problem defining two orbit segments, which can be implemented and solved within the software package \textsc{auto-07p} \cite{AUTO1,AUTO2}.

An orbit segment with an associated integration time $T$ is a solution of the time-rescaled differential equation
\begin{equation}\label{eqn:BVP}
	\frac{d \mathbf{x}}{dt} = T\vec{f}(\mathbf{x},\lambda_+).
\end{equation}
Here the function $f$ is given by (the real and imaginary parts of) the right-hand side of Eqs. (\ref{eqn:ODEs1})--(\ref{eqn:ODEs2}), $\mathbf{x}(t)=[b_{x}(t),b_{y}(t),\gamma(t)]$, and $\lambda_+$ is the continuation parameter. In this formulation, any orbit segment is represented over the unit time interval as 
\begin{equation}
	\mathbf{x}: [0,1] \rightarrow \mathbb{R}^3,
\end{equation}
and any boundary conditions imposed during a contination concern the begin point $\mathbf{x}(0)$ and/or the end $\mathbf{x}(1)$. The integration time $T$ is solved for during a computation and, in some formulations, it is allowed to vary. 

\subsection{Finding Homoclinic Orbits}

For the computation of homoclinic orbits to $\mathrm{N}_u$, we first compute two orbit segments: $\mathbf{x}_1$ with integration time $T_1$ in the one-dimensional unstable manifolds of $\mathrm{N}_u$ that ends in a suitable plane $\Sigma\subset\mathbb{R}^3$, and $\mathbf{x}_2$ with integration time $T_2$ in the two-dimensional stable manifolds of $\mathrm{N}_u$ that starts in $\Sigma$; for the purposes of the computations here, we choose $\Sigma = \{(b_x,b_y,\gamma)\in\mathbb{R}^3|\gamma = -0.4\}$. We subsequently continue both orbit segments in the system parameter $\lambda_+$ until their points $\mathbf{x}_1(1)$ and $\mathbf{x}_2(0)$ meet in $\Sigma$, meaning that $\mathbf{x}_1$ and $\mathbf{x}_2$ jointly form (a numerical approximation of) a homoclinic orbit; this is achieved by requiring that $\mathbf{x}_1(1)$ and $\mathbf{x}_2(1)$ lie (for any $\lambda_+$) along a fixed direction, given by the \emph{Lin vector}, which defines the \emph{Lin gap}; see Ref. \cite{krauskopf_lin_2008} for further technical details. The formulation of the respective boundary conditions below assumes knowledge of the eigenvalues and eigenvectors of $\mathrm{N}_u$ as a function of $\lambda_+$. 

\begin{figure}[t!]
	\centering
	\includegraphics[width=8.2cm]{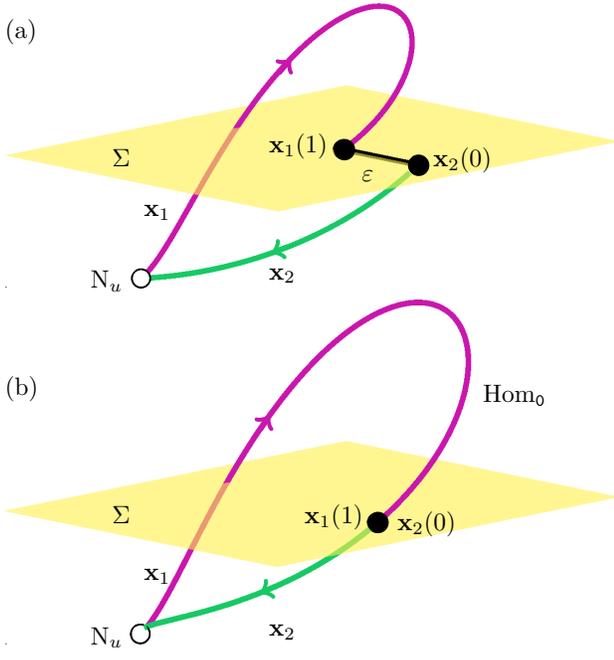}
	\caption{\label{fig:LinExample} Demonstration of Lin's method to compute a homoclinic orbit. Panel (a) shows the orbit segments $\mathbf{x}_{1}$ and $\mathbf{x}_{2}$ for $\lambda_{+} = 1.5$ after Steps 1A and 1B with a nonzero Lin gap $\varepsilon$ between them. Panel (b) shows them forming the homoclinic orbit $\mathrm{Hom}^1 = \mathrm{Hom}_{\zero}$ for $\lambda_{+} = 1.53286$ from Fig. \ref{fig:orbits}(a), found in Step 2 when $\varepsilon=0$. Other parameters are as in Fig. \ref{fig:OneParBifDiag}.}
\end{figure}

\begin{enumerate}[leftmargin=20mm]
	\item[\textit{Step 1A}:] Compute, via continuation in $T$, the unstable manifold orbit segment $\mathbf{x}_1\subset W^{u}_-(\mathrm{N}_u)$. The boundary conditions are
	\begin{align}
		\mathbf{x}_{1}(0) &= \mathrm{N}_{u} + \delta_1\vec{y}_{u}, \label{eqn:unstab1} \\
		\mathbf{x}_{1}(1) &\in \Sigma. \label{eqn:stab1}
	\end{align}
	Condition (\ref{eqn:unstab1}) represents a first-order approximation of the one-dimensional unstable manifold of $\mathrm{N}_u$, where $\vec{y}_u$ is its unstable eigenvector and $\delta_1$ is sufficiently small. Initially, $\mathbf{x}_{1}(1)$ does not lie in $\Sigma$ and the continuation in $T$ stops when it does and (\ref{eqn:stab1}) is satisfied.\\
	
	\item[\textit{Step 1B}:] Compute, also via continuation in $T$, a stable manifold orbit segment $\mathbf{x}_2\subset W^{s}(\mathrm{N}_u)$. The boundary conditions are
	\begin{align}
		\hspace{12mm}
		\mathbf{x}_{2}(0) &\in \Sigma, \label{eqn:unstab2} \\
		\mathbf{x}_{2}(1) &= \mathrm{N}_{u} + \delta_{2}\left[\vec{y}_{s}\cos(\Phi) + \vec{y}_{ss}\sin(\Phi) \right] \label{eqn:stab2}.
	\end{align}
	Condition (\ref{eqn:stab2}) represents the first-order approximation of the  two-dimensional unstable manifold of $\mathrm{N}_u$, spanned by the eigenvectors $\vec{y}_{s}$ and $\vec{y}_{ss}$ of its two real stable eigenvalues; here the chosen angle $\Phi$ determines the direction of approach of the orbit segment to $\mathrm{N}_u$ \cite{krauskopf_computing_2007}. Again, the continuation in $T$ stops when condition (\ref{eqn:unstab2}) is satisfied.\\
	
	\item[\textit{Step 2}:] Close the Lin gap via continuation in $\lambda_{+}$. We define the Lin vector as 
	\begin{equation*}
		\vec{v} \equiv \mathbf{x}_{2}(0) - \mathbf{x}_{1}(1)
	\end{equation*}
	at the beginning of this step and, importantly, then keep $\vec{v}$ fixed throughout (even though $\mathbf{x}_{1}(0)$ and $\mathbf{x}_{2}(1)$ subsequently change with $\lambda_{+}$). We then continue both orbit segments $\mathbf{x}_{1}$ and $\mathbf{x}_{2}$ subject to (\ref{eqn:unstab1}), (\ref{eqn:stab1}), (\ref{eqn:stab2}), and the additional boundary condition
	\begin{equation}\label{eqn:LinDir}
		\mathbf{x}_{2}(0) = \mathbf{x}_{1}(1) + \varepsilon \vec{v},
	\end{equation}
	which ensures that the Lin direction remains constant and defines the Lin gap $\varepsilon$. Here $\lambda_+$, $T_1$, $T_2$, $\Phi$, and $\varepsilon$ are free to vary, and the continuation stops when $\varepsilon = 0$ is detected, which means that $\mathbf{x}_{2}(0) = \mathbf{x}_{1}(1)$ and a homoclinic orbit has been found.
	
\end{enumerate}

Figure \ref{fig:LinExample} illustrates Lin's method with the example of finding the homoclinic orbit $\mathrm{Hom}^1 = \mathrm{Hom}_{\zero}$ from Fig. \ref{fig:orbits}(a). Figure \ref{fig:LinExample}(a) shows the situation for $\lambda_{+} = 1.5$ after Steps 1A and 1B, when the two orbit segments $\mathbf{x}_{1}$ from  $\mathrm{N}_u$ to the section $\Sigma$ and $\mathbf{x}_{2}$ from $\Sigma$ to $\mathrm{N}_u$ have been found. The Lin gap $\varepsilon$ between the two endpoints $\mathbf{x}_{1}(1), \mathbf{x}_{2}(0) \in \Sigma$ is initially nonzero, and it is then closed during Step 2. This zero of $\varepsilon$ is found for $\lambda_{+} = 1.53286$, and Fig. \ref{fig:LinExample}(b) shows that then $\mathbf{x}_{1}(1) = \mathbf{x}_{2}(0)$ and the two orbit segments $\mathbf{x}_{1}$ and $\mathbf{x}_{2}$ jointly form the homoclinic orbit $\mathrm{Hom}_{\zero}$. 

Further intersection points $\mathbf{x}_{1}(1) \in \Sigma$ of $W^{u}_-(\mathrm{N}_u)$ with the section $\Sigma$ can be found with Step 1A by further continuation in $T$. Each such point defines the associated Lin gap with $\mathbf{x}_{2}(0) \in \Sigma$, which is then closed in Step 2. In this way, we computed the other homoclinic orbits $\mathrm{Hom}_{\zero\one}$, $\mathrm{Hom}_{\zero\one\zero}$, etc. Note that the symmetric counterparts of any of these homoclinic orbits, formed by $W^{u}_+(\mathrm{N}_u)$, are given by the images of $\mathbf{x}_{1}$ and $\mathbf{x}_{2}$ under rotation by $\pi$ about the $\gamma$-axis.

\subsection{Numerical Continuation of EtoP Connections}

The numerical implementation of Lin's method can also be used to find an EtoP connection; in fact, it was developed for this purpose in Ref. \cite{krauskopf_lin_2008}. We use it here to find EtoP connections from 
the equilibrium $\mathrm{N}_u$ to either of the periodic orbits $\mathrm{SRO}_u$ when they are of saddle type. To this end, we consider an orbit segment $\mathbf{x}_{2}$ that now lies in the two-dimensional stable manifold $W^{s}(\mathrm{SRO}_u)$. Such an orbit segment is defined by the boundary condition 
\begin{equation}\label{eqn:BCEtoP}
	\mathbf{x}_{2}(1) = p_{0} + \delta_2\vec{w}_{s}, 
\end{equation}
where $p_{0}$ is a point on the periodic orbit ($\mathrm{SRO}_u$ with either $b_x< 0$ or $b_x> 0$, depending on the connection sought) and $\vec{w}_{s}$ is the vector at $p_{0}$ of the stable linear Floquet bundle of the periodic orbit. By replacing boundary condition (\ref{eqn:stab2}) with (\ref{eqn:BCEtoP}) in Step 1B, an orbit $\mathbf{x}_{2} \in W^{s}(\mathrm{SRO}_u)$ with $\mathbf{x}_{2}(0) \in \Sigma$ is found by continuation in $T$; here $\delta_2$ is a chosen small distance of $\mathbf{x}_{2}(1)$ from the periodic orbit. 

The thus found point $\mathbf{x}_{2}(0)$ defines the Lin gap $\varepsilon$ with $\mathbf{x}_{1}(1)$ as before and Step 2 is then used to close it, where the distance $\delta_2$ is free to vary (instead of the angle $\Phi$ as for homoclinic orbits). We remark that the periodic orbit as well as its stable bundle must be computed, and continued in Step 2 when the parameter $\lambda_{+}$ changes. The technical details of the required extended boundary value problem setup can be found in Ref. \cite{krauskopf_lin_2008}. As for homoclinic orbits, we considered further intersection points $\mathbf{x}_{1}(1) \in \Sigma$ of $W^{u}_-(\mathrm{N}_u)$ found in Step 1A to find the more complicated EtoP connections, such as $\mathrm{EtoP}_{\zero\one\one\bar{\zero}}$; their symmetric counterparts are again given by rotation of $\mathbf{x}_{1}$ and $\mathbf{x}_{2}$ by $\pi$ about the $\gamma$-axis.

\bibliography{SGKP_PoleFlip.bib}

\end{document}